\documentclass[conference]{IEEEtran}
\usepackage{cite}
\usepackage{amsmath,amssymb,amsfonts}
\usepackage{algorithmic}
\usepackage{graphicx}
\usepackage{textcomp}

\usepackage[utf8]{inputenc}
\usepackage{subcaption}
\usepackage{fancyhdr}

\begin{document}

\title{Towards Digital Twin Oriented Modelling of Complex Networked Systems and Their Dynamics: A Comprehensive Survey}


\author{\uppercase{Jiaqi Wen}, \uppercase{Bogdan Gabrys and Katarzyna Musial}.}

\markboth
{Author \headeretal: Preparation of Papers for IEEE TRANSACTIONS and JOURNALS}
{Author \headeretal: Preparation of Papers for IEEE TRANSACTIONS and JOURNALS}


\maketitle
\thispagestyle{empty}
\pagestyle{plain}

\begin{abstract}
This paper aims to provide a comprehensive critical overview on how entities and their interactions in Complex Networked Systems (CNS) are modelled across disciplines as they approach their ultimate goal of creating a Digital Twin (DT) that perfectly matches the reality. We propose a new framework to conceptually compare diverse existing modelling paradigms from different perspectives and create unified assessment criteria to assess their respective capabilities of reaching such an ultimate goal. Using the proposed criteria, we also appraise how far the reviewed current state-of-the-art approaches are from the idealised DTs. We also identify and propose potential directions and ways of building a DT-orientated CNS based on the convergence and integration of CNS and DT utilising a variety of cross-disciplinary techniques.
\end{abstract}




\section{Introduction}
\label{sec:introduction}

A complex network can be seen as a universal concept used for representation and analyses of complex systems. Given the growing interest in real complex systems and a fast development of modelling techniques, the complex networked system (CNS) area has become a highly cross-disciplinary field that involves multiple modelling approaches with various research aims posed and achieved over the years.

There is a considerable literature about complex networks and various researchers have published several surveys reviewing and exploring the topic from different perspectives and application areas. Those include works on complex networks and their applications covering multiple application areas \cite{IEEEexample:costa2011analyzing} or orientated towards specific topics such as networks of cryptocurrency transactions \cite{IEEEexample:liu2021knowledge}, vehicular networks \cite{IEEEexample:celes2021mobility}, internet of things \cite{IEEEexample:batool2017modeling} and networks of short written text \cite{IEEEexample:amancio2015probing}.

When it comes to the models of complex networks in the context of their structure, there is also a body of work surveying a variety of network topologies~\cite{IEEEexample:amancio2015probing,IEEEexample:costa2007characterization} and dynamics. The networks dynamics can be either considered as: (i) dynamic processes over networks, that involve surveys on spreading processes like epidemic processes, information spreading processes \cite{IEEEexample:brodka2020interacting,IEEEexample:nowzari2016analysis}, or (ii) dynamic networks with evolving structures and features \cite{IEEEexample:skarding2020foundations,IEEEexample:rossetti2018community}.

When it comes to modelling techniques that have been applied to complex networked systems, there are also surveys that review certain types of modelling approaches for complex networks, including Graph Neural Network \cite{IEEEexample:skarding2020foundations}, game theory \cite{IEEEexample:manshaei2013game} or non-parametric Bayesian modeling \cite{IEEEexample:schmidt2013nonparametric}.

Complex networked systems are modelled with the goal of accurate reflection of reality with the aim of simulation, prediction and/or control. Over the years, proposed CNSs models have become more and more accurate with more realistic networks topologies, characteristics and evolving dynamics modelled. Increasingly, they can capture features of real world scenarios and behave like their twins. Researchers have already focused on the studies of Digital Twining of real systems across disciplines, which involves a wide range of applications but so far complex networks area has been marginalised in this development space.

Therefore, the convergence of Digital Twinning and Complex Networked Systems emerges as an exciting research focus with a potential to address some of the outstanding modelling and representational challenges ultimately leading to the establishment of a DT-orientated CNS area as its main goal.

Multiple modelling approaches have been applied to answer various questions about CNSs, and there is a need to review and explore where, how and why complex networked systems are modelled across many different disciplines. However, most of the surveys only account for certain types of complex networked system from a specific perspective or a single discipline. Therefore, to fill in this gap, we review research on complex networks from a holistic view while trying to deal with the questions concerning diverse modelling paradigms and their distance to the idealised DTs. We devise a framework that enables to compare various modelling paradigms and evaluate their respective distances to DTs, which involves answering the following four fundamental questions: (1) What is the aim of the modelling?; (2) How to represent information about a system in a form of a network?; (3) How to model the dynamics in a networked system?; and (4) How do we approach the ultimate goal of building a CNS that models the reality at a Digital Twin level?

Hence, while answering the above questions, this critical survey aims to integrate and overview the current, relevant state-of-the-art from multiple disciplines and inform future research directions and foci for this new multidisciplinary area.

Our paper is organised as follows. Section~\ref{aim} illustrates and discusses different modelling aims across different disciplines and summarises their research foci. This is followed by two sections concerned with complex networked systems with section~\ref{how1} reviewing different network representational and modelling paradigms and section~\ref{quadrant} focusing on modelling dynamics in networked systems. In Section~\ref{performance} a new framework is proposed for evaluating models of complex network systems given the ultimate goal of achieving DTs faithully representing the reality. The research gaps and future research directions are also identified and discussed in this section. Finally, the Conclusions are provided in the last section.

\section{Model's Aims}
\label{section2}
There is always a need to determine the aim of developing a new model at the start of the modelling process.  The choice of modelling paradigms, which can be understood as the approaches employed to model real state or dynamics of the system, is largely dependent on the model's aims determined based on the research questions. The model's aims include a wide range of functions like link prediction, network detection, mimicking of real systems, etc. while they are developed given different disciplinary focuses such as epidemiology, sociology, microeconomics, etc. We categorise the modelling paradigms based on their concrete model's aims, which helps us to understand what is expected from modelling complex network systems and serves as the basis for further discussions of detailed modelling paradigms. In this section, we also set an ultimate goal of modelling complex networked systems, having in mind various model's aims, and this goal is to create a Digital Twin (DT) of real world system.

\subsection{Prerequisites of setting and fulfilling model's aims}

Selecting the right modelling paradigm together with the availability of the needed input data are the prerequisites of setting and fulfilling model's aims. Given varying complexity of reality which can only be partially observed and modelled, there are no concrete standards or measures to ensure and prove the implementation of these prerequisites. Thus the model is always built on the strong assumptions of appropriate variable and model selection. The discrepancy between these assumptions and the truth may result in incapability of predicting dramatic changes due to partially observable input space \cite{IEEEexample:degenhardt2019evaluation} or overfitting problem that is typical for overly complex models \cite{IEEEexample:lever2016points}.
Therefore, researchers start to evaluate and reduce their distance to these prerequisites by considering observability.

\textbf{Observability,} in structural way, is about the ability of reconstructing the state of a system from a limited set of measured variables in finite time \cite{IEEEexample:aguirre2018structural}, while from the perspective of dynamics, observability is related to a deeper understanding about how the selected variables interact and evolve to change the states of the system. The structural observability is dependent on the available data set while the dynamical observability is analysed based on a model of dynamics, which is inevitably wrong to some extent but may be useful either in reflecting the reality or diagnosing an input space that is partially observable. Linear time-invariant system proposed by \cite{IEEEexample:kailath1980linear} is one typical modelling paradigm widely used in the studies on observability of dynamical systems \cite{IEEEexample:leitold2017controllability}.

In complex networked systems (CNS), structural observability determines the number of nodes to be measured to render a network observable, while dynamical observability considers node dynamics and coupling when selecting the best driver (sensor) node to modulate (observe) the whole network activity \cite{IEEEexample:aguirre2018structural}.
The design and analysis of CNS for each model's aim involves the observability of topology \cite{IEEEexample:leitold2017controllability}, variables for coupling nodes \cite{IEEEexample:sendina2019observability} and node dynamics \cite{IEEEexample:iudice2019node}. These structural observability and dynamical observability are closely related to variable and model selection, and serve as the prerequisites of representing and modelling reality under each model's aim in different application scenarios.

The measures of observability vary depending on the linearity of systems. Linear networks without symmetries have been well studied research objects with respect to the topic of observability, where the observability matrix based on the dynamic model for a linear (time-invariant) system proposed by \cite{IEEEexample:kailath1980linear} is widely used. Nonlinearity of dynamical networks have been recently considered in the studies on dynamical observability. For example, \cite{IEEEexample:whalen2015observability} quantify the observability and controllability of nonlinear networks with explicit symmetries that shows the connection between symmetries and nonlinear measures of observability and controllability. \cite{IEEEexample:letellier2018nonlinear} propose a nonlinear graph-based theory for dynamical network observability from the Jacobian matrix of the governing equations of nonlinear systems.

\subsection{Model's aims in different research fields}
\label{aim}
The model's aims can be categorised as: (i) specific and (ii) abstract goals. Specific model's aims, such as community discovery, link prediction, anomaly detection, synchronization and controllability of networked systems, focus on specific external tasks for observable research object (i.e. CNS in our case) with measurable model outputs. Abstract model's aims, like topological feature analysis and the mimics of real life systems, approach the inner rules of real dynamics, analyse observables and simulate unobservables for further research on specific model's aims. Examples of different CNS models' aims together with the relevant references are shown in Table \ref{tab1}:

\begin{table}
\caption{Different goals of building CNS models}
\label{tab1}
\setlength{\tabcolsep}{3pt}
\begin{tabular}{|p{75pt}|p{150pt}|}
\hline
 Model's aim & References\\
\hline
 Community discovery  &\cite{IEEEexample:rossetti2018community}, \cite{IEEEexample:alvari2014community}, \cite{IEEEexample:qin2018adaptive}, \cite{IEEEexample:jin2015combined}, \cite{IEEEexample:rosvall2010mapping}, \cite{IEEEexample:nguyen2011adaptive}, \cite{IEEEexample:braun2015dynamic}, \cite{IEEEexample:macon2012community} \\
       Link prediction  &  \cite{IEEEexample:gao2017community}, \cite{IEEEexample:lu2011link}, \cite{IEEEexample:martinez2016survey}, \cite{IEEEexample:liben2007link}, \cite{IEEEexample:wahid2019predict}, \cite{IEEEexample:clauset2008hierarchical}, \cite{IEEEexample:fouss2006experimental}, \cite{IEEEexample:nayyeri2021trans4e}, \cite{IEEEexample:lu2009similarity} \\
       Node classification &
       \cite{IEEEexample:rossi2007measurability}, \cite{IEEEexample:ashraf2019simulation}, \cite{IEEEexample:lei2019gcn}, \cite{IEEEexample:wang2020nodeaug}\\
       Synchronization  & \cite{IEEEexample:arenas2008synchronization}, \cite{IEEEexample:dorfler2012synchronization}, \cite{IEEEexample:dorfler2013synchronization}, \cite{IEEEexample:bullmore2012economy}, \cite{IEEEexample:tang2013distributed}, \cite{IEEEexample:hammond2007pathological}, \cite{IEEEexample:wong2008optimizing}, \cite{IEEEexample:wickramasinghe2013spatially}, \cite{IEEEexample:sohn2017small}\\
       Controllability & \cite{IEEEexample:rossetti2018community}, \cite{IEEEexample:alvari2014community}, \cite{IEEEexample:cazabet2011simulate}, \cite{IEEEexample:rosvall2010mapping}, \cite{IEEEexample:nguyen2011adaptive}, \cite{IEEEexample:braun2015dynamic}, \cite{IEEEexample:macon2012community} \\
       Anomaly detection & \cite{IEEEexample:ranshous2015anomaly}, \cite{IEEEexample:bindu2017discovering}, \cite{IEEEexample:yu2018netwalk}, \cite{IEEEexample:kendrick2018change}, \cite{IEEEexample:thottan2003anomaly}, \cite{IEEEexample:rajasegarar2008anomaly}, \cite{IEEEexample:bhuyan2013network}, \cite{IEEEexample:cancer2008comprehensive}, \cite{IEEEexample:jia2018pattern} \\
       Topological analysis& \cite{IEEEexample:amancio2015probing}, \cite{IEEEexample:i2004patterns}, \cite{IEEEexample:backes2009complex}, \cite{IEEEexample:bullmore2012economy}, \cite{IEEEexample:zhang2018convergence}, \cite{IEEEexample:kaviani2020influence}, \cite{IEEEexample:ganesh2005effect}, \cite{IEEEexample:sohn2017small}\\
       Mimics of reality&  \cite{IEEEexample:krol2015propagation}, \cite{IEEEexample:nowzari2016analysis}, \cite{IEEEexample:keeling2005networks}, \cite{IEEEexample:amblard2004role}, \cite{IEEEexample:cho2012identification}, \cite{IEEEexample:ye2018open}, \cite{IEEEexample:sohn2017small}, \cite{IEEEexample:atzori2012social}, \cite{IEEEexample:cancer2008comprehensive}, \cite{IEEEexample:soh2010weighted}, \cite{IEEEexample:morais2021learning}. \\
\hline
\end{tabular}
\end{table}

\textbf{Community discovery} aims to decompose complex networks into meaningful sub-networks that better describe local phenomena \cite{IEEEexample:rossetti2018community}. The local phenomena refers to a set of entities that share some closely correlated sets of actions with the other entities of the community \cite{IEEEexample:coscia2011classification,IEEEexample:rossetti2018community}. This has been explored and discussed in a wide range of applications, including the detection of community structure hidden in real social networks \cite{IEEEexample:alvari2014community,IEEEexample:qin2018adaptive,IEEEexample:jin2015combined}, collaboration network analysis like detecting citation patterns \cite{IEEEexample:rosvall2010mapping}, improving routing of telecommunication network \cite{IEEEexample:nguyen2011adaptive}, reconfiguration of the brain network \cite{IEEEexample:braun2015dynamic} or political affiliation\cite{IEEEexample:macon2012community}.

\textbf{Link prediction} aims to infer the behaviour of the network link formation process by predicting missed or future relationships based on observed links and the attributes of both nodes and relationships \cite{IEEEexample:lu2011link,IEEEexample:martinez2016survey}. Link prediction involves questions of dealing with missing links or link labels of networks and predicting links in changing networks, including social networks \cite{IEEEexample:liben2007link,IEEEexample:wahid2019predict,IEEEexample:gao2017community}, food webs \cite{IEEEexample:clauset2008hierarchical}, networks in collaborative recommendation tasks\cite{IEEEexample:fouss2006experimental}, knowledge graphs \cite{IEEEexample:nayyeri2021trans4e} and biochemical networks of protein interaction \cite{IEEEexample:lu2009similarity} and metabolism \cite{IEEEexample:clauset2008hierarchical}. 

\textbf{Node classification} aims to provide a labeling for unlabeled nodes in a network composed of partially labelled nodes and edges \cite{IEEEexample:rossi2007measurability}. Node classification, as an important way to explore node features and links, has been widely studied in social networks \cite{IEEEexample:rossi2007measurability,IEEEexample:ashraf2019simulation}, citation networks \cite{IEEEexample:lei2019gcn} and co-author networks \cite{IEEEexample:wang2020nodeaug}.

\textbf{Synchronisation} of complex networks implies that the states of two or more interacting nodes in a network with different initial conditions gradually approach each other and finally reach the same state  \cite{IEEEexample:arenas2008synchronization}. The applications of synchronisation in complex networks range from the stability of power grids \cite{IEEEexample:dorfler2012synchronization,IEEEexample:dorfler2013synchronization}, controllablity of neuronal networks \cite{IEEEexample:bullmore2012economy,IEEEexample:tang2013distributed,IEEEexample:hammond2007pathological}, optimising timetables for transportation \cite{IEEEexample:wong2008optimizing} to the synchronization patterns affected by network topology in chemical systems \cite{IEEEexample:wickramasinghe2013spatially} and IoT systems \cite{IEEEexample:sohn2017small}.

\textbf{Controllability} of networks represents the ability of controlling the networks, which is independent of the way that the outputs are formed, while its related concept of observability, depends only on the outputs but not on the inputs \cite{IEEEexample:kalman1963mathematical}. 
The studies on controllability are often combined with observability, which range from social networks \cite{IEEEexample:cremonini2017controllability}, protein interaction networks \cite{IEEEexample:wuchty2014controllability}, brain networks \cite{IEEEexample:karrer2020practical} to transportation networks \cite{IEEEexample:rinaldi2018controllability}.

\textbf{Anomaly detection} of networks is about finding objects, relationships, or points in time that are unlike the rest \cite{IEEEexample:ranshous2015anomaly}. There are many studies on anomaly detection in various application scenarios, ranging from anomaly detection in social networks \cite{IEEEexample:bindu2017discovering,IEEEexample:yu2018netwalk,IEEEexample:kendrick2018change}, public health \cite{IEEEexample:yu2018netwalk}, IP networks \cite{IEEEexample:thottan2003anomaly}, wireless sensor networks \cite{IEEEexample:rajasegarar2008anomaly} to intrusion of networks \cite{IEEEexample:bhuyan2013network}. Similarly to the principles of anomaly detection, pattern recognition is also used for diagnostic analysis \cite{IEEEexample:cancer2008comprehensive} and heterogeneous component detection in knowledge networks \cite{IEEEexample:jia2018pattern}.

\textbf{Topological feature analysis} is a very popular model's aim in both real-life networks and artificial networks. This is a research topic that probes network topological features from real data for applications across disciplines based on network-based models, such as the probed topological features of text network for language organisation \cite{IEEEexample:amancio2015probing,IEEEexample:i2004patterns}, probed boundary features within the small world complex networks for image classification \cite{IEEEexample:backes2009complex} and organisation of brain network \cite{IEEEexample:bullmore2012economy}. Local network topological features like three-node motif \cite{IEEEexample:juszczyszyn2008temporal}, directed closure \cite{IEEEexample:jia2021directed} and quadrangle structure \cite{IEEEexample:jia2021measuring} are researched through topological feature analysis of social networks. There are also discussions of artificial network topological effects on network based models, which involves studies of topological effects on P2P trading in financial market \cite{IEEEexample:zhang2018convergence}, artificial neural networks in computation tasks \cite{IEEEexample:kaviani2020influence}, epidemic spreading \cite{IEEEexample:ganesh2005effect} and enhancing synchronization of IoT systems\cite{IEEEexample:sohn2017small}.

\textbf{Mimics of reality} helps to deal with questions about analyses of dynamics over networks or features of networks \cite{IEEEexample:krol2015propagation}. There are studies of spreading processes in artificial networks including epidemic dynamics \cite{IEEEexample:nowzari2016analysis,IEEEexample:keeling2005networks}, opinion dynamics \cite{IEEEexample:amblard2004role,IEEEexample:cho2012identification} and meme diffusion \cite{IEEEexample:ye2018open}. There are also studies that simplify the real complex systems as data-driven networks to assist further analyses, which involves applications ranging from Digital Twins of IoT systems \cite{IEEEexample:sohn2017small,IEEEexample:atzori2012social}, image information representation \cite{IEEEexample:cancer2008comprehensive}, mimics of transportation systems \cite{IEEEexample:soh2010weighted} and the representation of human-object interactions via networks \cite{IEEEexample:morais2021learning}.

\subsection{Digital Twin: an ultimate goal}
Researchers focus on the studies of twining real systems across many disciplines and those efforts have already resulted in a development of a field on its own known as Digital Twins (DT). DT serves as an "almighty" paradigm of mimics across spatial and temporal scales. It has also grown to become an ultimate goal of modelling complex networked systems due to its reality-friendly nature, integration of model functions and the wide range of applications.

Digital Twin is a virtual extension of reality, which not only allows to compare
current conditions with historical data to provide meaningful information to assist in decision-making, but also enables forecasting and feedback of eventualities that have never happened before \cite{IEEEexample:jans2020digital}.
Researchers have defined DTs from different perspectives across the application scenarios. In a fully digitalized product life cycle, DT is a comprehensive virtual product model with the features of real-time monitoring, simulation and forecasting \cite{IEEEexample:HAAG201864}. For mechanical and cyber-physical systems, DT is a linked collection of digital artefacts that evolves with the real system along the whole life cycle and integrates currently available knowledge with the purpose of describing behaviour and deriving solutions for the real system \cite{IEEEexample:boschert2018next}.

DTs have three elementary components repeatedly emphasised: the digital (virtual part), the real physical product and the connection between them \cite{IEEEexample:grieves2014digital}, while there are also other imperative components added with the accumulation of practice, including data, service, machine learning, and DT Performance evaluation  \cite{IEEEexample:tao2018digital,IEEEexample:tao2019digital,IEEEexample:sharma2020digital}.
DTs also resemble a series of models with integrated functions like simulation, optimization and data analytics \cite{IEEEexample:ivanov2019digital} and features of real-time processing and continuous updates \cite{IEEEexample:madni2019leveraging}. This makes DTs impossible to be replaced by any single tool and ideal modelling paradigms for health monitoring \cite{IEEEexample:tao2018digitall} , planning of manufacturing \cite{IEEEexample:uhlemann2017digital}, management of smart city \cite{IEEEexample:du2020cognition}, accurate healthcare \cite{IEEEexample:liu2019novel} and anomaly detection \cite{IEEEexample:castellani2020real} within a wide range of complex systems, including complex networked systems like  IoT systems \cite{IEEEexample:alam2017c2ps,IEEEexample:singh2014survey} and blockchain-encapsulated systems  \cite{IEEEexample:he2018blockchain,IEEEexample:hao2020towards}.

The above mentioned components, integrated functions and universal application of DTs differentiate them from any other simulation tool or modelling paradigms by emphasising the properties of real-time data acquisition of observations and feedback, and self-evolution through continuous machine learning analysis. They contribute to DTs' status as a powerful tool for the mimicking of a series of realities and an ultimate goal for modelling complex networked systems across disciplines.

Modelling complex networked systems using a Digital Twin paradigm has a potential to build a universal model that can be adapted to fulfil multiple, different models' aims discussed in this section. But before this can be attempted, we need to review and assess how the modelling of complex networks, their dynamics as well as dynamics on those networks is approached currently and this is the focus of the following sections.

\section{How to represent information in a form of a network}
\label{how1}
This question involves two important issues in building complex networked systems, which are: (1) what types of networks are needed in modelling certain phenomena including the ways of building the required network topology with appropriate complexity using real data and simulations, and (2) how to obtain these networks from different perspectives of obtaining and processing data to realise such complexity. To achieve a faithful representation of a network that preserves as much information as needed in the modelling of networked systems (see section \ref{how1}) for certain models' aims (see \ref{aim}), one needs to collect and process observable data and information about the network structures and associated dynamics with modelling paradigms that minimise the information loss in the process of a network generation.

\subsection{How to represent networked data}
\label{NT}
According to \cite{IEEEexample:breitbart2000topology} network topology is a representation of the physical connections that exist among entities in a communication network. This definition can be and has been easily expanded across disciplines as network topology describes how the entities of any type relate to each other in any type of network. The establishment of network topology is a crucial step for modelling complex networked system, and its effect on the dynamics has been a popular research focus.

\subsubsection{Complexity dimensions}
\begin{figure*}[t!]
\centering
\vspace{0.1cm}
\setlength{\abovecaptionskip}{0.3cm}
\setlength{\belowcaptionskip}{-0.3cm}
  \includegraphics[width=5.5 in]{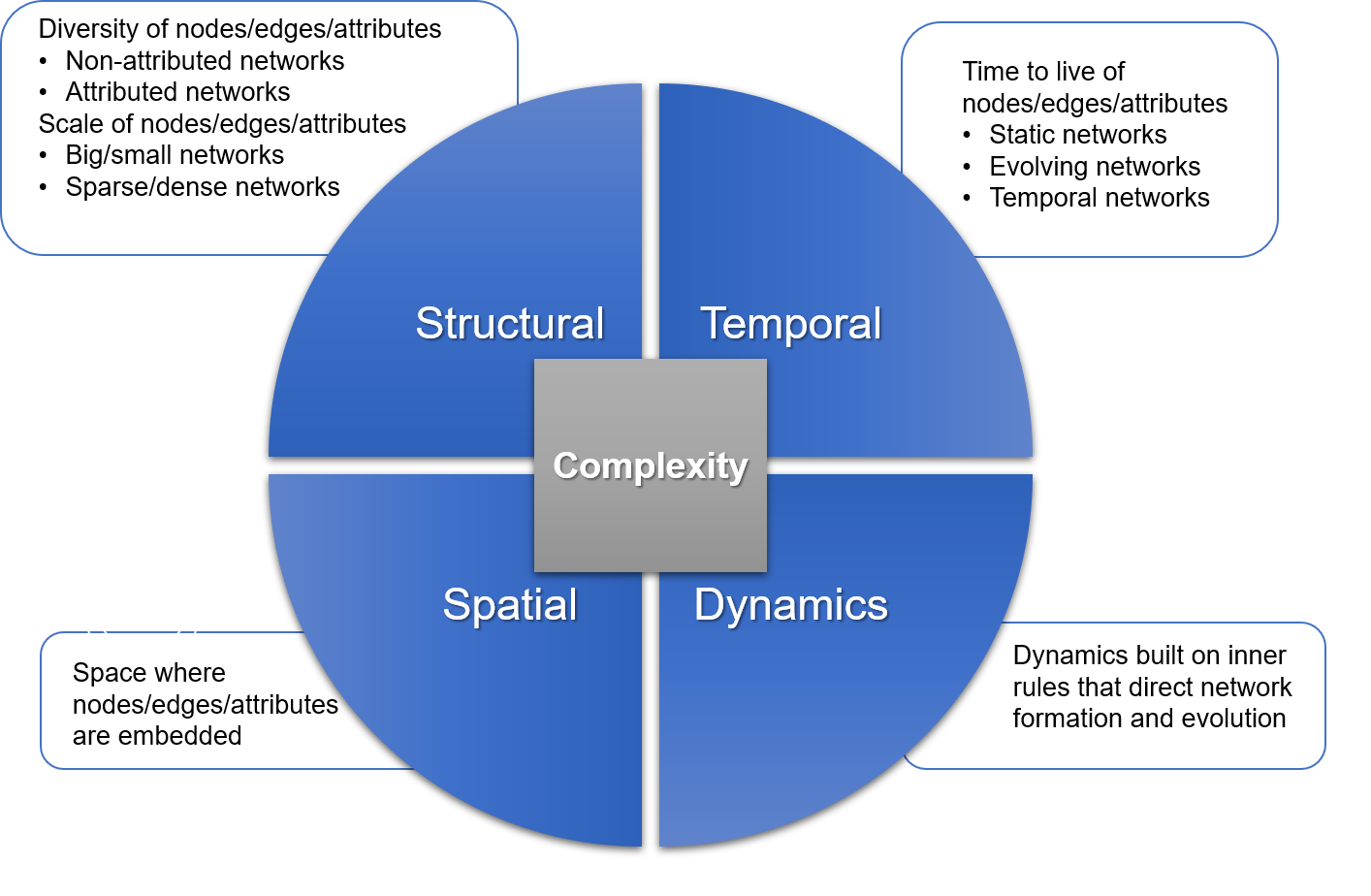}
\caption{Four dimensions of complexity related to a network's topology.}
\label{Fig0}      
\end{figure*}

Network topologies vary in complexity when they represent networked information that can vary in data availability and modelling necessity. Complexity of the topology results from different types of nodes, edges and their attributes. As shown in Figure \ref{Fig0}, we propose to describe this complexity alongside and using the following four dimensions: (i) a structural dimension connected to the scale and diversity of the topology, (ii) a temporal dimension concerned with time-to-live of different components of the network, (iii) a spatial dimension connected with the space in which the topology can be embedded, and (iv) a dynamics dimension connected to the topology's exerted or encapsulated dynamics. The complexity in structural, temporal and spatial dimensions describes the necessary reality required to be represented via networks, while the dynamics complexity depends on the models selected to explore the complex reality. These modelling paradigms are shown in Figure \ref{Fig0}. Given available networked data set that is observable from the perspective of the above mentioned four complexity dimensions, in topological feature analysis, only structural information about networks need to be represented and analysed based on a model that reveals the inner rules of network formation, while for prediction of link formation over time, temporal complexity is further incorporated and a complex model that performs well in such an external task is employed.

\paragraph{Structural complexity}
Structural complexity of the scale and diversity of nodes, edges and their attributes involves discussions of non-attributed and attributed networks in terms of their diversity, as well as small/big networks and sparse/dense networks in terms of their scale, where structural complexity increases with more detailed information needed to be represented.

Research studies usually start from the exploration of non-attributed networks that are built only with homogenous nodes and edges. Such networks represent simplified real world scenarios and are thoroughly studied in terms of their topological features, like the analysis of potential energy landscape with both a small-world and scale-
free character \cite{IEEEexample:doye2002network} or the conduction of specific model's aim e.g. community discovery which only uses network topology to find partitions \cite{IEEEexample:tyler2003email}.

Attributed networks can better represent real-world networked interactions and information as they introduce the auxiliary information via node or edge attributes \cite{IEEEexample:shah2016edgecentric,IEEEexample:hou2020rosane}.
The node attributes describe the features of nodes within interactions or relations, while the edge attributes capture information about how the adjacent nodes interact with others in the networks \cite{IEEEexample:shah2016edgecentric}. These attributes vary with application scenarios. Taking online social networks as an example, nodes represent users and are attributed with user profiles \cite{IEEEexample:huang2021hop}, while edges represent online relationships and have attributes like nature of relation, direction, intensity and durability \cite{IEEEexample:musial2013social}.
Many researchers study the structural complexity of node-attributed networks having in mind such modelling aims as community discovery \cite{IEEEexample:jin2018robust,IEEEexample:qin2018adaptive}, link prediction \cite{IEEEexample:li2018streaming}, anomaly detection \cite{IEEEexample:huang2021hop}, controllability \cite{IEEEexample:li2016connectivity}, topological feature analysis and mimicking of reality \cite{IEEEexample:xia2020dynamic}.
However, few researches focus on the representation and modelling of generalised edge-attributed networks. Edge-attributes of such networks within most researches typically take the numerical or categorical forms \cite{IEEEexample:shah2016edgecentric,IEEEexample:liu2020semi}, where directions \cite{IEEEexample:liu2020jordan} and edge weights \cite{IEEEexample:duong2012effective,IEEEexample:nair2018networked} are often studied. Compared with the large number of studies on node-attributed networks with various model's aims, only a small number of studies focuses on community discovery \cite{IEEEexample:kang2018edge} and anomaly detection \cite{IEEEexample:shah2016edgecentric} of edge-attributed networks.

The structural complexity also increases with the requirements of representing more information within large-scale networks and the difficulty of processing networks with sparse edge information.
Structural complexity of large-scale networks with thousands and millions of nodes results from a complicated and higher-order inner structure \cite{IEEEexample:bedru2020big}, which are common in DTs like city IoT \cite{IEEEexample:xia2020dynamic} and DT of manufacturing with big data \cite{IEEEexample:fei2018digital}.
Structural complexity of sparse networks, given fewer edges, lies in their restrictions of attributes processing \cite{IEEEexample:hou2020rosane,IEEEexample:chunaev2020community} and optimal modelling \cite{IEEEexample:krzakala2013spectral}. There is an even more complex case for large sparse networks where both complicated large-scale inner structure and problematic sparse edges are involved \cite{IEEEexample:krzakala2013spectral}.

\paragraph{Temporal complexity}
Temporal complexity of networks increases when more temporal information about nodes, edges and their attributes can be captured and modelled meaning that less information is lost. Networks can be conceptually described as static, edge-weighted,
evolving and temporal \cite{IEEEexample:rossetti2018community,IEEEexample:skarding2020foundations} as they are transforming to be instantaneous and approaching the state of continuity without temporal aggregation.

The basic modelling of real-world phenomena and systems using CNS starts with a static network topology where nodes and edges are fixed and they are assumed to be "frozen" in time. Such an assumption greatly simplifies the modelling process but fails to capture the evolving features of real-world systems.

The attempts of accounting for temporal information in the network modelling process have gained attention as they usually improve the model performance. For example, social network analysis initially viewed networks as static rather than changing over time~\cite{IEEEexample:marsden1982social,IEEEexample:perry2003social}. As the field developed, social interactions started being represented using temporal networks to capture dynamics and instantaneous character of the contacts \cite{IEEEexample:nair2018networked,IEEEexample:kefalas2018recommendations}. In terms of biomolecular networks, studies on protein-protein networks first employed static data-driven networks to represent and analyse the protein-protein interaction \cite{IEEEexample:li2010computational} until the dynamic protein-protein networks have been found to benefit the study on the molecular systems of protein complexes \cite{IEEEexample:chen2014identifying}. The analyses of transportation networks also initially used static networks \cite{IEEEexample:mandl1980evaluation} and then shifted to dynamic management of transportation systems for better model performance \cite{IEEEexample:wu2001computational}.
However, the transition from the static networks under the most stringent assumptions to the temporal networks that result in smaller loss of information can not be achieved overnight, while the scale of involving time dimension also differs from case to case.

Some of the researches focus on edge-weighted networks with temporal weights in the analysis of social relations \cite{IEEEexample:nair2018networked} and the mobility in wireless networks~\cite{IEEEexample:duong2012effective}. Other researches focus on evolving networks where network topology changes slowly over time so that its instantaneous snapshot yields a well defined network \cite{IEEEexample:skarding2020foundations}. Some evolving networks just represent more stable relations, rather than instantaneous interactions between the nodes, which can be captured with durable edges like citations \cite{IEEEexample:bruner2010appraisal} and friendships \cite{IEEEexample:musial2013social,IEEEexample:gao2016hybrid}, while some researches use evolving networks built with snapshots to aggregate the temporal information of interactions within a time window for a more stable representation in the analysis of instantaneous features~\cite{IEEEexample:nguyen2011adaptive,IEEEexample:rosvall2010mapping,IEEEexample:cazabet2011simulate,IEEEexample:gao2016hybrid}. There are also studies on temporal networks that have non-trivial topology changes and can not be represented via instantaneous snapshots. They preserve all the temporal information and build networks in a more faithful way, such as instantaneous contacts of communication via e-mail, text messages, or phone calls with temporal edges of networks \cite{IEEEexample:holme2012temporal}.
\paragraph{Spatial complexity}
Spatial complexity involves discussion of spatial networks, which are proposed and defined as networks where nodes are located in a space equipped with a metric of the usual euclidean distance \cite{IEEEexample:barthelemy2011spatial}. For example, spatial networks that represent urban street patterns can be built with a metric of distance, which is measured not just in topological terms (steps), but in properly spatial terms (meters, miles)
\cite{IEEEexample:crucitti2006centrality}.
Considerable applications of spatial networks involve modelling of human activities that take place on a spatial matrix obtained from largely three types of transportation networks including matter (streets, roads, highways, railways, airport networks), energy (the power grid) and information (Internet, telephone networks) \cite{IEEEexample:buhl2006topological,IEEEexample:cardillo2006structural}.

Researchers have been characterising and understanding the structure and the evolution of networks under the impact of spatial constraints  \cite{IEEEexample:barthelemy2011spatial}, where topology alone does not contain all the information to understand the dynamics of networks and the spatial information is needed. Some social-spatio networks consider the interplay of social networks and locations, such as social interactions under the impact of transportation \cite{IEEEexample:goetzke2015social}, employment outcomes in a labour market driven by social contacts under an explicit geographic structure \cite{IEEEexample:topa2019social} and the social properties of Twitter users’ networks with the spatial proximity of the networks \cite{IEEEexample:stephens2015follow}. Traffic under the constraints of transportation networks also involves studies on commuter flows constrained in large transportation networks \cite{IEEEexample:ren2014predicting} and planning of unobstructed paths in traffic-aware spatial networks \cite{IEEEexample:shang2015planning}.

\begin{figure}[t!]
\centering
\vspace{0.1cm}  
\setlength{\abovecaptionskip}{0.3cm}
\setlength{\belowcaptionskip}{-0.3cm}
  \includegraphics[width=3 in]{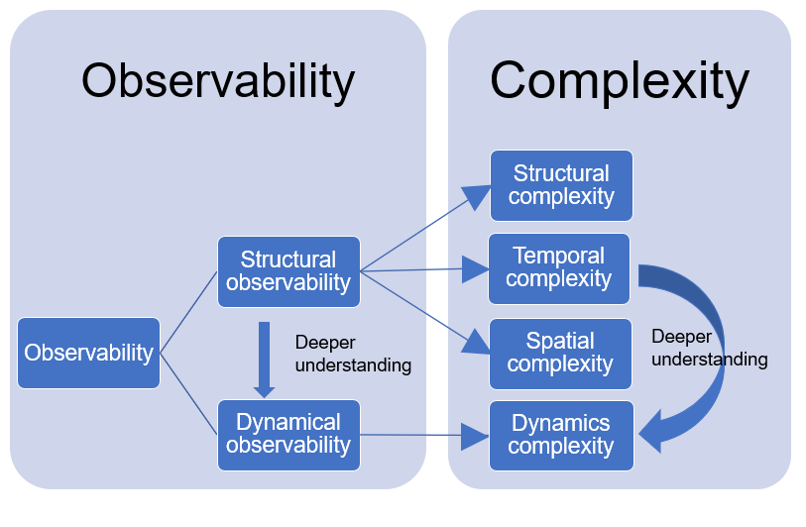}
\caption{Observability and complexity with corresponding relations}
\label{Fig1}
\end{figure}

\paragraph{Dynamics complexity}
Dynamics complexity is about the dynamics of networks that enables to develop a deeper understanding of temporal complexity based on structural and spatial complexity by investigating the rules of network evolution via simulation or modelling. Temporal and dynamics complexity, which are respectively derived from structural and dynamical observability, represent varying degrees of reality. The relations between observability and complexity dimensions are shown in Figure \ref{Fig1}.


For dynamics built on inner rules that direct  network formation and evolution, the dynamics complexity increases as networks are generated with less human involvement based on more realistic rules ranging from statistical relations \cite{IEEEexample:solomonoff1951connectivity,IEEEexample:gilbert1959random} to realistic principles like homophily \cite{IEEEexample:mcpherson2001birds,IEEEexample:asikainen2020cumulative}. This dynamics can either be exerted dynamics that controls the network changes to a desired state of CNS \cite{IEEEexample:verhoeven2020controlling,IEEEexample:shi2020evaluating,IEEEexample:block2020social}, or encapsulated dynamics that motivates the network formation and evolution with higher degrees of automation \cite{IEEEexample:mcpherson2001birds,IEEEexample:ashraf2019simulation,IEEEexample:asikainen2020cumulative}. 
For example, exerted dynamics used for social network intervention can control each step of network evolution by changing attributes of nodes that are identified based on different man-made strategies \cite{IEEEexample:shi2020evaluating}. Encapsulated dynamics mainly focuses on the edge formation mechanisms, e.g. based on preferential attachment principles \cite{IEEEexample:ashraf2019simulation,IEEEexample:asikainen2020cumulative} which are typical examples of highly-automated self-evolution without involvement of human decision.

\paragraph{The synergy effects of complexity dimensions}
The synergy effects of complexity dimensions describe the "1+1>2" effect of the combined complexities from different dimensions, which are more complex but closer to reality.
Spatial temporal networks are typical examples for combined complexity of temporal and spatial dimension, which are proposed for a more faithful representation of reality with the influence of space on constraining the structure of temporal networks considered. Some researches employ temporal networks to capture and process temporal information under consecutive frames, while they construct spatial networks to extract certain static features. Such spatial-temporal networks have been widely used in computer vision like facial expression \cite{IEEEexample:zhang2017facial}, video-based person reidentification \cite{IEEEexample:xu2017jointly} and identification of human-object interaction \cite{IEEEexample:morais2021learning}.
Some researches introduce temporal information to spatial networks. Taking recommendation task as an example, \cite{IEEEexample:kefalas2018recommendations} build spatial temporal networks with temporal edge-weights by incorporating time dimension to user-location graphs and using sessions that capture the co-locations among two or more users during a time window.

It is clear that different complexity dimensions are intertwined. They influence and build on each other and this is one of the challenges that need to be considered when modelling of complex networked systems is attempted. Research has been conducted in each of the complexity dimensions but building overarching framework that would enable to flexibly adjust a level of complexity of each of the dimensions and simulate various what if scenarios is still an outstanding challenge. With the recent developments in the Digital Twin space, modelling CNS using DT paradigm is a promising way forward.

\subsubsection{Data-driven vs simulation-based networks}
\label{datasimulation}
\begin{figure}[t!]
\centering
\vspace{0.1cm}
\setlength{\abovecaptionskip}{0.3cm}
\setlength{\belowcaptionskip}{-0.3cm}
  \includegraphics[width=3 in]{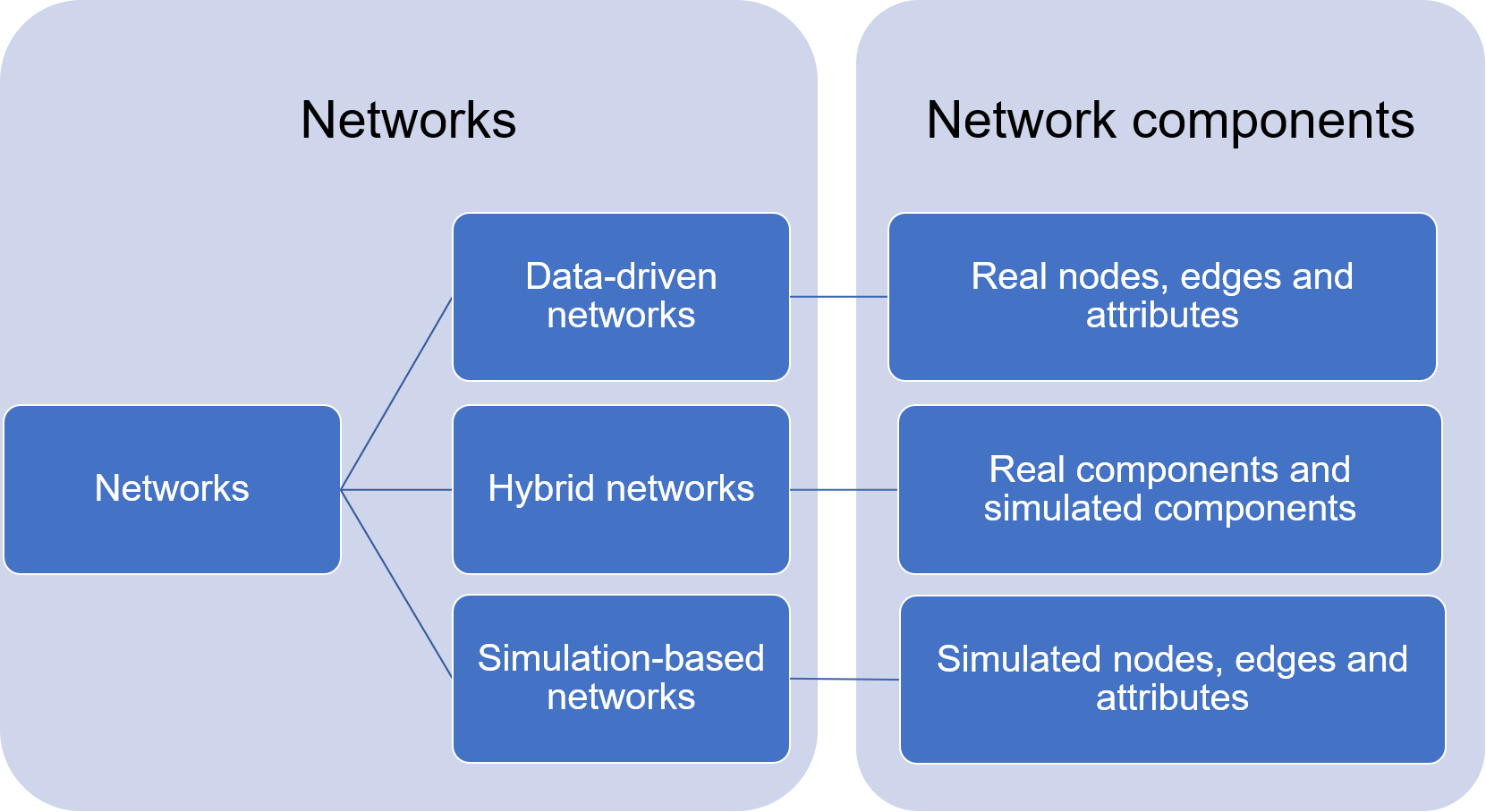}
\caption{Different types of networks and their components.}
\label{Fig2}       
\end{figure}

Researchers have made a lot of effort to faithfully represent the information from real world systems by developing variety of modelling approaches. This involves data-driven networks, simulation-based networks and networks that are built by combining these two approaches. These networks are featured with varying degrees of complexity. Components of different network types are shown in Figure \ref{Fig2}.


Data-driven networks built on rich real data sets may not necessarily capture all the information as they may be confined by relatively simple model's aims under strict assumptions, like the ignorance of temporal information for temporal networks analysed in a "static" time scale \cite{IEEEexample:marsden1982social}. As researchers explore the real data sets with more complex modelling aims and advanced techniques, data-driven networks gradually approach reality as more complexity is allowed to be introduced to network topology with more relaxed assumptions. For example, the analysis of social networks started from statistical analysis of static networks and their topology \cite{IEEEexample:marsden1982social} and then evolved into a more complex modelling aims such as community discovery in networks that are increasingly more and more complex, starting from static \cite{IEEEexample:cai2005mining} to evolving \cite{IEEEexample:alvari2014community,IEEEexample:wang2016autonomous}, attributed \cite{IEEEexample:chunaev2020community} and both evolving and attributed networks \cite{IEEEexample:rossetti2018community}. 

When real data is not available, simulated (synthetic) networks can be generated and used to analyse various phenomena~\cite{IEEEexample:brodka2020interacting}. These simulated networks enable to model network phenomena with different levels of complexity. The statistics-based simulations based on predetermined network statistics involve classical examples like the Barabasi-Albert model for the scale-free network~\cite{IEEEexample:barabasi1999emergence}, the Watts-Strogatz small-world model for the small-world network~\cite{IEEEexample:watts1998collective} or the Erdos-Renyi model for the random graph network~\cite{IEEEexample:solomonoff1951connectivity,IEEEexample:gilbert1959random}. The simulation-based approaches increase their complexity and flexibility as more complex rules governing creation of structure and its dynamics are introduced. The capability of generating networks with distinctive features enabled the similutaion-based approaches to become a universal tool for topological feature analysis. They also help with assessing the impact of network topology on dynamic processes, involving researches like synchronization of IoT systems using networks ranging from scale-free to small world models \cite{IEEEexample:sohn2017small}.

Another type of simulation-based networks, the principle-based simulations, are built according to different connection principles like homophily \cite{IEEEexample:mcpherson2001birds,IEEEexample:asikainen2020cumulative}, triadic structure \cite{IEEEexample:asikainen2020cumulative}, geographic proximity \cite{IEEEexample:block2020social,IEEEexample:zhang2021modeling}. These networks have typically higher degrees of temporal and dynamics complexity than statistics-based network simulations as they self-evolve with highly autonomous and interpretable edge formation process and generate temporal networked information.
However, few researches involve principle-based network simulations with various edge attributes \cite{IEEEexample:ashraf2019simulation} or network simulations embedded in spatial dimension \cite{IEEEexample:kim2020location}. They are more complex but closer to reality, which calls for future research on network simulations with the combination of different levels of complexity from structural, temporal, spatial and dynamics space.

There are also networks built using combination of real data and simulations. They are featured with the mix of different levels of complexity either captured from real data or represented via simulation. Such networks consist of real attributed nodes and connections between the nodes are created using network simulation dynamics. For example,  \cite{IEEEexample:shi2020evaluating} build social networks by extracting necessary information about nodes' attributes from real database and simulate the edges using scale-free networks based on a network density from referenced literature. Some researches simulate missing edges for real networks, where the single imputation methods including null-tie imputation and reconstruction summarised by \cite{IEEEexample:krause2020missing} are typical examples. The complexity of networks built using this hybrid approach increases as more data-driven features are captured by having the network simulation dynamics trained to fit observable real network components. For example, as an improvement of single imputation, researches propose a multiple imputation method that fits an exponential random graph model (ERGM) to the real data and simulates missing ties via inference \cite{IEEEexample:krause2020missing}. 

\subsection{How to generate networks using different modelling paradigms}
\label{obtain}
Networks, no matter to what extent real data or simulation is used, are created to faithfully represent information about a given system in preparation for further analyses. Depending on the modelling goal, networks used for analysis will differ with respect to the four introduced above complexity dimensions and will have varying settings of nodes, edges and attributes.
There are several modelling paradigms that are used to obtain desired networks and those include rule--based, agent--based and event--based approaches that focus on fundamental generative process from a local perspective of network formation, as well as basic graph--based, probabilistic graph--based and network--embedding based approaches that aim at condense network representation from the global perspective. These modelling paradigms, of local or global level, differ in ways of observing and processing networks but converge to a faithful representation of reality that aims at minimising the information loss between reality and the model. Observed local interactions lead to global emerging characteristic behaviours observed and analysed via graphs, while graphs lose less information as the four complexity dimensions are introduced via local level observations.


\subsubsection{A Local view}
\label{how1.1}
Modelling paradigms of networks from a local view focus on mechanisms ruling the network formation. They take local level observations as a starting point and explore how they construct the characteristic CNS at a global level, corresponding to the discussion of observability considering the structural, temporal, spatial and dynamics complexity. The rule--based, agent--based and event--based approaches at a local level are introduced below with more information observed or simulated for the four complexity dimensions.

\paragraph{Rule--based paradigm}
Rule--based paradigm generates networks under explicit dependence laws from predetermined assumptions or rules detected from real world cases, which not only focuses on network topology, but also investigates dynamics of network structures via simplified rules of edge formation and evolution.

Rule--based paradigms are controlled by a limited number of parameters according to a rule-based mathematical function, where randomness is introduced via certain probability distributions involved in edge formation or variable changes, like the scale-free network generation via a scale-free power-law distribution \cite{IEEEexample:fortunato2006scale}, edge formation with a bias probability dependent on the similarity of node attributes \cite{IEEEexample:asikainen2020cumulative}, and the changes of node property according to a uniform distribution \cite{IEEEexample:ashraf2019simulation}. For rule-based paradigms, there is inevitable information loss resulting from the divergence of rules and reality, as the rules greatly simplify the partially observable and complex real world scenarios. Researchers have made much effort to bridge this gap, involving studies on rule-based dynamics transforming from simulated \cite{IEEEexample:barabasi1999emergence,IEEEexample:watts1998collective,IEEEexample:solomonoff1951connectivity,IEEEexample:gilbert1959random} to trained to better fit real networks \cite{IEEEexample:asikainen2020cumulative}, as well as rules evolving from statistics-based \cite{IEEEexample:broido2019scale,IEEEexample:hunter2008goodness,IEEEexample:musial2013creation,IEEEexample:krause2020missing,IEEEexample:godoy2021inference} to principle-based \cite{IEEEexample:zhang2010finite,IEEEexample:asikainen2020cumulative}. They include temporal complexity by incorporating temporal changes of network topology into the rule--based network generation process, and sometimes, to better characterise network generation, introduce the impact of node attributes with increasing structural complexity.

Rule--based paradigm started from simulated dynamics of networks either based on statistical rules \cite{IEEEexample:barabasi1999emergence,IEEEexample:watts1998collective,IEEEexample:solomonoff1951connectivity,IEEEexample:gilbert1959random} or principles \cite{IEEEexample:mcpherson2001birds,IEEEexample:zhang2010modeling,IEEEexample:asikainen2020cumulative} (as is mentioned in section \ref{datasimulation}.), where they can be tuned to approach reality via seeking optimal parameters for the rule-based mathematical functions to fit real data and make inference \cite{IEEEexample:asikainen2020cumulative}. For example, the ERGM involved in multiple imputation of networks can fit real data and simulate missing ties via inference \cite{IEEEexample:krause2020missing}, where the involved dynamics represent data-driven features to some extent, but neither preserve rich information about node/edge attributes nor explore topological information other than typical interactive patterns of ERGM rules.

Statistics-based network generators are able to fit real networks via statistical inference based on an explicit likelihood function, like scale-free model \cite{IEEEexample:broido2019scale}, ERGM \cite{IEEEexample:hunter2008goodness,IEEEexample:krause2020missing} and geometric branching growth model \cite{IEEEexample:godoy2021inference}. As statistics-based dynamics on typical interactive patterns can hardly represent diverse real-world networks, principle-based network generators have been studied with their flexible and adaptable design of principles, which fit the statistics of real networks via likelihood-free inference such as approximated Bayesian computation \cite{IEEEexample:asikainen2020cumulative}. For example, scale-free structure is empirically rare in social networks \cite{IEEEexample:broido2019scale}, while a principle-based simulator based on cumulative effects of triadic closure and homophily is able to reveal social network dynamics \cite{IEEEexample:asikainen2020cumulative}.

\paragraph{Event--based paradigm}
Event--based paradigm refers to the network representations with two elementary components: nodes and their local pairwise interactions referred to as events \cite{IEEEexample:arora2017action} or nodes of events and their logical relationships \cite{IEEEexample:choi2013modeling,IEEEexample:macedo2015context,IEEEexample:pham2015general}, which involves event-based network representation and analysis \cite{IEEEexample:macedo2015context,IEEEexample:pham2015general,IEEEexample:liu2012event} as well as stochastic point processes for events that can perform network inferential tasks \cite{IEEEexample:fox2016modeling}.

Event--based paradigm started from a simple network representation that captures richer heterogeneous information of interactions related to events with increasing structural complexity. As a typical example, event-based social networks (EBSNs) are enabled to further capture and use the information of offline social interactions, in addition to the online social relationships included in conventional online social networks \cite{IEEEexample:liu2012event}. The enriched information observed about events assist further analysis and modelling tasks. For example, the consideration of both online and offline interactions for EBSNs provides adequate information for global level analysis and modelling via graph-based models \cite{IEEEexample:pham2015general} and improves the prediction power for event recommendation tasks \cite{IEEEexample:liu2012event}.

Then it comes to stochastic point processes that are enabled to make inference about nodes \cite{IEEEexample:fox2016modeling} or edges (events) \cite{IEEEexample:passino2021mutually} via estimating dependencies between events or dependencies between events and latent space models with increasing dynamics complexity. Taking Hawkes process as an example, \cite{IEEEexample:fox2016modeling} employ a self-exciting point process on the edge to perform an identity-inference task, considering the effect of available observations of geographically distributed interactions (edges). \cite{IEEEexample:passino2021mutually} apply mutually exciting point process on the edge that includes the effect of node-specific latent vectors to assess the significance of previously unobserved connections for anomaly detection tasks.

\paragraph{Agent--based paradigm}
Agent--based paradigm applied in the context of complex networked system refers to interactive structures that are typically composed based on three basic components:
agents, interaction rules, and space (this could be geo-space or some other abstract space) \cite{IEEEexample:epstein2006generative}, which enables more degrees of freedom in building networks from rich data or simulations as they can select more detailed information about interactions from microscopic perspective of agents.

Agent--based paradigms vary in degrees of information loss with different dimensions and degrees of complexity required for various research goals.
The interactive structure can either be static with neighbour sets determined once and for all, or dynamic with evolution along time depending on model assumptions \cite{IEEEexample:bargigli2014interaction}, while the interaction rules may either be simulation-based under certain constraints of space \cite{IEEEexample:zhang2018convergence,IEEEexample:wang2019simulation,IEEEexample:andrei2014agent} or data-driven based on realistic scenarios \cite{IEEEexample:venkatramanan2018using,IEEEexample:long2018data,IEEEexample:wang2021multi}.

The information about interactions of agents is initially represented in a fixed network topology, where decisions of agents are affected by network structures over which they interact. Such networks serve as an environment and a constraint for agents' behaviours \cite{IEEEexample:bargigli2014interaction}. Most static networks involved in the researches on agent-based models are simulation-based, focusing on the network effect on interactions in agent-based systems like trading behaviours in double auction \cite{IEEEexample:zhang2018convergence,IEEEexample:zhang2018influence} and tax compliance and evasion \cite{IEEEexample:andrei2014agent}. Recently, these agent-based networks approach reality by introducing real data with features of time and space. \cite{IEEEexample:venkatramanan2018using} proposed a data-driven agent-based model for forecasting emerging infectious diseases, where data-driven networks are built with spatial information and social contact of realistic synthetic population. \cite{IEEEexample:long2018data} also proposed an agent-based computational model under a data-driven decision-making framework for supply chain networks given their complicated micro structures, macro emergencies and dynamic evolution.

Multi-agent systems are also important simulation tools for modelling evolving networked systems with multiple agents interacting under certain constraints. \cite{IEEEexample:cazabet2011simulate} proposed a multi-agent system that replays the evolution of a network and reproduces the rise and fall of communities with the strength of adapting to real-time, changing problems. \cite{IEEEexample:nedic2010constrained} research on the constrained concensus and optimisation of multi agent networks, where  multiple agents align their estimates with a particular value over a network with time-varying connectivity in different local constraint sets.

To better capture the structural patterns and instantaneous dynamics of networks, event-driven models (also named as activity-driven models) have been proposed with an activity potential, a time invariant function characterising the agents’ interactions and encoding the instantaneous time description of the network dynamics \cite{IEEEexample:perra2012activity}. These paradigms include the rich information observed or simulated from microscopic views of agents and model their connections activated by an event trigger, involving observations varying from binary interactions \cite{IEEEexample:perra2012activity} to simplicial complexes \cite{IEEEexample:petri2018simplicial}, as well as applications ranging from event-based consensus \cite{IEEEexample:meng2013event} to contagion problems \cite{IEEEexample:petri2018simplicial,IEEEexample:ogura2019optimal}.

\subsubsection{A Global view}
\label{how1.2}
Modelling paradigms of networks from a global view aim at representing high-dimensional and heterogeneous networked information in a way that can be easily analysed. More information can be observed at a global level as more information about interactions is captured via local methods, which involves discussion on how global performance can change through controlling and modifying those interactions, corresponding to the research aim about controllability. The basic graph--based, probabilistic graph--based and network embedding--based methods at a global level are introduced below with increasing complexity as they are able to preserve more observable information in network representation and modelling.

\paragraph{Basic graph--based paradigm}
Basic graph--based paradigm is based on the graph theory and it can be seen as a set of selection principles for microscopic laws of behaviour in network science \cite{IEEEexample:iniguez2020bridging} which typically involves a simplified graph representation and analysis of networks just concerned with nodes and their connections, e.g. \cite{IEEEexample:aittokallio2006graph,IEEEexample:george2018graph}.

Graph theory began when, in 1735, Leonhard Euler presented the first mathematical demonstration based on geometry of position to solve the seven bridges of K\"oningsberg puzzle
\cite{IEEEexample:euler1741solutio,IEEEexample:alexanderson2006cover}. Graph theory focuses on providing rigorous proofs for graph properties, such as graph enumeration, coloring, and covering \cite{IEEEexample:bondy1976graph,IEEEexample:iniguez2020bridging}, while the evolution of random graphs motivated graph theory to generate a new branch of network science for a separate direction: quantifying the structure and dynamics of real-world complex systems \cite{IEEEexample:iniguez2020bridging}.

Graph-based paradigm simply represents networks as basic graphs composed of nodes and connections, enabling readily available graph analysis but taking the cost of information loss to varying degrees especially in terms of structural complexity ~\cite{IEEEexample:aittokallio2006graph,IEEEexample:djidjev2011graph,IEEEexample:george2018graph}. For example, from an accumulation of experimental data on biomolecules, graph-based models for cell biology build the graph only with cellular components (nodes) and their interactions (edges), which allows for network topology analyses using graph-theoretical concepts but lose information other than the graph-structure \cite{IEEEexample:aittokallio2006graph}.

Graphs are one of the widely studied data structures in computer science and discrete mathematics \cite{IEEEexample:washio2003state}, while the graph-based models are also widely applied in the modelling of CNS across disciplines, such as analyses of graphic characteristics for networks in cell biology~\cite{IEEEexample:aittokallio2006graph}, anomaly detection of computer networks using protocal graphs~\cite{IEEEexample:djidjev2011graph}, graph representation of vulnerability relations for industry IoT network~\cite{IEEEexample:george2018graph}.

\paragraph{Probabilistic graph--based paradigm}
Probabilistic graph--based paradigm models networks with uncertainties on the relationships between nodes \cite{IEEEexample:hassanlou2013probabilistic}, which has two elementary components: a graph that defines the network structure and a set of local functions whose product is the joint probability of this compact representation \cite{IEEEexample:sucar2015probabilistic}. The network representation with probabilistic graph--based paradigm is typically flexible in directed or undirected, static or dynamic dimensions, each corresponding to varying degrees of structural complexity and temporal complexity. The exploration of the local functions for dependence rules or cause-effect relationships enables the inference and learning of real networks in sophisticated models
\cite{IEEEexample:anzai2012pattern}.


The probabilistic graphs can either be undirected with symmetric relations like conditional random fields \cite{IEEEexample:lafferty2001conditional} and Markov networks \cite{IEEEexample:taskar2007relational,IEEEexample:jin2020modmrf}, or directed with cause-effect relationships between the nodes, such as sigmoid belief networks \cite{IEEEexample:sutskever2008deep}, Bayesian networks \cite{IEEEexample:chen2012good} and hidden Markov model \cite{IEEEexample:beal2002infinite}. The usage of directed or undirected graphs depends on the application scenarios. For example, Markov networks, as a typical undirected graph, represent relational dependencies without the hindrance of acyclicity constraint and thus are well suited for discriminative training \cite{IEEEexample:taskar2007relational}. They have been widely used in argumentation tasks like finding labellings or probabilistic inference tasks by deciding credulous and sceptical acceptance \cite{IEEEexample:potyka2020abstract}. While Bayesian networks, as a directed acyclic graph with explicit cause-effect assumptions for interactions \cite{IEEEexample:chen2012good}, can handle problems including missing data, prediction and data over-fitting \cite{IEEEexample:heckerman2008tutorial}. Some researchers also employ Bayesian networks in DTs due to efficient linear computation \cite{IEEEexample:li2017dynamic,IEEEexample:alam2017c2ps,IEEEexample:yu2021digital}, while nonparametric Bayesian networks are more flexible in capturing time varying features with computation efficiency \cite{IEEEexample:schmidt2013nonparametric,IEEEexample:orbanz2014bayesian}.

Probabilistic graph--based paradigm can either be static or dynamic based on whether the probabilistic graphs represent variables at a time point or across different times \cite{IEEEexample:sucar2015probabilistic}, such as the Bayesian networks \cite{IEEEexample:chen2012good} versus the dynamic Bayesian networks \cite{IEEEexample:li2017dynamic}. They are transforming from static to dynamic with the trend of modelling networks with more temporal complexity. For example, continuous time Bayesian networks\cite{IEEEexample:boudali2006continuous} and continuous time Markov networks \cite{IEEEexample:anderson2011continuous} are proposed to capture changeable variables given the evolving features of structured stochastic processes and the graphical structure of Markov networks and Bayesian networks.

Probabilistic graph--based paradigms are able to conduct inference that helps in answering different probabilistic queries based on the model and some evidence as well as learning that estimates the graph structure and parameters of probabilistic graph--based paradigms' local functions \cite{IEEEexample:sucar2015probabilistic}. As exact inference is often intractable, researchers tend to use approximation algorithms to find distributions that are close to the correct posterior distribution \cite{IEEEexample:frey2005comparison}, like Gibbs sampling \cite{IEEEexample:plummer2003jags} and belief propagation \cite{IEEEexample:sudderth2010nonparametric}. While maximum likelihood estimation methods and the expectation maximization (EM) algorithms \cite{IEEEexample:gracca2007expectation} are respectively employed for learning problems with or without hidden variables. However, probabilistic graph--based paradigms have difficulty representing, inferring and learning high-dimensional, heterogeneous networks, which calls for combined application of network embedding methods.

\paragraph{Network embedding--based paradigm}
Network embedding--based paradigm aims at network construction and network inference via embedding node information in the network into low-dimensional space \cite{IEEEexample:cui2018survey}, which is featured with a concatenation of an encoder and a decoder \cite{IEEEexample:budka2013molecular,IEEEexample:hamilton2017representation,IEEEexample:skarding2020foundations}.

Network embedding starts from dimensionality reduction techniques that are also applicable in scenarios other than networks, which include stochastic multidimensional scaling \cite{IEEEexample:rajawat2017stochastic}, isometric mapping (ISOMAP) \cite{IEEEexample:li2012novel}, principle component analysis (PCA) \cite{IEEEexample:bollen2009principal}, linear discriminant analysis (LDA) \cite{IEEEexample:paul2014decision}, stochastic neighbor embedding (SNE) \cite{IEEEexample:yang2014optimization} and
t-distributed stochastics neighbor embedding (t-SNE) \cite{IEEEexample:olivon2018metgem}. There are also basic models that merely focus on network embedding tasks, which, as summarised by \cite{IEEEexample:wang2020brief}, are respectively built upon the skip-gram models \cite{IEEEexample:guthrie2006closer} and matrix factorization models \cite{IEEEexample:singh2008unified}. These models are able to encode network information into a low-dimensional space, but do not focus on decoding information to reconstruct networks or perform inferential tasks.

To preserve more information in the modelling process and perform inferences, deep learning (DL) techniques have been widely utilised to embed highly diverse, heterogeneous, high-dimensional network information into a low-dimensional latent space. DL-based network embedding, also referred to as a network representation learning, is able to make inferences and assist network analytic tasks including node classification, link prediction, clustering, recommendation, similarity search and visualization, involving unsupervised and semi-supervised learning methods \cite{IEEEexample:zhang2018network}. As it is categorised by \cite{IEEEexample:cai2018comprehensive}, DL based graph embedding methods are either random walk based like Skip-Gram based deep learning models \cite{IEEEexample:perozzi2014deepwalk,IEEEexample:grover2016node2vec,IEEEexample:yanardag2015deep}, or without random walk but directly utilising DL methods on a whole graph or its proximity matrix via autoencoder \cite{IEEEexample:gong2019network}, deep neural networks \cite{IEEEexample:cao2016deep} and graph convolutional networks \cite{IEEEexample:chen2020multi}.

Graph Neural Networks (GNN) refer to a widely used DL-based embedding method that encode graph structures via a neural networks architecture, which is able to decrease information loss by aggregating features of neighbouring nodes together \cite{IEEEexample:skarding2020foundations}. This motivates the combination of GNN with other models, which is typically composed of an encoder, a generative model and a decoder. For example, graph Bayesian networks \cite{IEEEexample:sun2020framework} and graph Markov Neural Networks \cite{IEEEexample:qu2019gmnn,IEEEexample:bacciu2018contextual} can be employed to infer graph parameters (statistical relational learning) based on encoder/decoder - graph convolutional neural networks = and generative model = the involved PGM. The GNN combined with ordinary differential equation (ODE) is able to infer the latent states of irregular observations (encoding), learn the state transition in latent space via generative model = ODE, and make predictions continuously (decoding) \cite{IEEEexample:huang2021coupled}.

\subsubsection{An Overall view}
Three trends can be discovered if we look through the modelling paradigms of generating networks from either local or global view: (i) increasing complexity of network topology and dynamics, (ii) decreasing interpretability with the increase of complexity, (iii) models' aims transforming from abstract to specific. These trends are shown in Figure \ref{Fig3}:


\begin{figure*}[t!]
\centering
\vspace{0.1cm}  
\setlength{\abovecaptionskip}{0.3cm}
\setlength{\belowcaptionskip}{-0.3cm}
  \includegraphics[width=6 in]{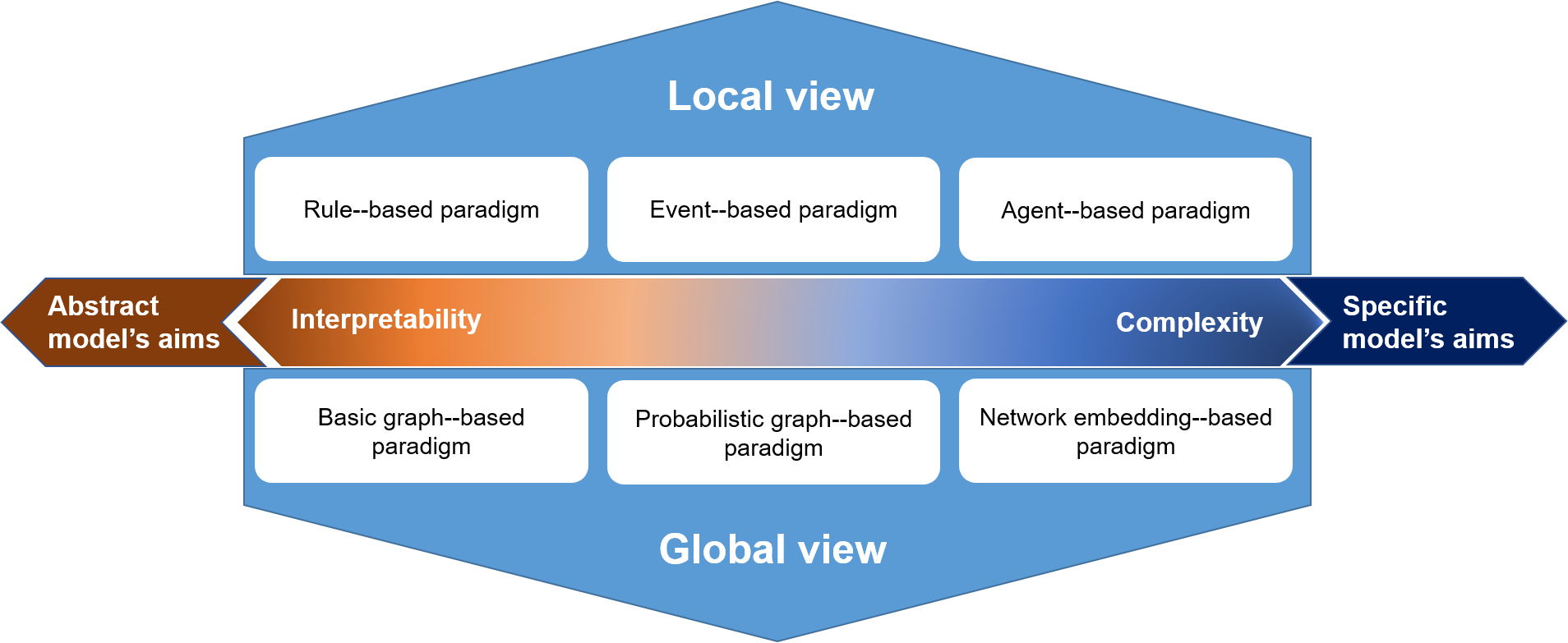}
\caption{An overall view of modelling paradigms}
\label{Fig3}       
\end{figure*}

\textbf{Complexity} of networks increases when they are represented and generated more faithfully with less information loss.

Models at a local level focus on the network formation process. Rule--based paradigms are usually statistics-based and generate static networks that are just composed of fixed nodes and edges \cite{IEEEexample:fortunato2006scale}. They recently turned to be principle-based while allowing for the introduction of node attributes with increasing structural complexity and edge addition over time with increasing temporal complexity \cite{IEEEexample:ashraf2019simulation,IEEEexample:asikainen2020cumulative}. Given the above complexity dimensions, rule--based paradigms generate networks under the simplest rules of edge addition with lowest level of dynamics complexity. To incorporate more information about attributes and temporal changes of network topology including edge addition and removal, event--based paradigms based on more complex rules of local pairwise interactions, such as the stochastic point processes, are employed with increasing dynamics complexity \cite{IEEEexample:fox2016modeling}. Agent--based paradigms introduce more complex interaction rules to account for more detailed information of agents (nodes), including their attributes and various actions that result in edge addition or removal over time \cite{IEEEexample:cazabet2011simulate}.

Models at a global level aim at a condense representation of high-dimensional and heterogeneous networked information. Basic graph--based paradigms focus on a simplified graph representation of networks that are just about nodes and their connections \cite{IEEEexample:aittokallio2006graph}. Probabilistic graph--based paradigms further incorporate information of network components with increasing structural and temporal complexity via the introduction of node attributes, edge directions or addition of edges over time \cite{IEEEexample:sucar2015probabilistic,IEEEexample:martinez2016survey}. These models, with increasing dynamics complexity, also enable the inference and learning of real networks via modelling the uncertainties on the relationships between nodes \cite{IEEEexample:hassanlou2013probabilistic}. Network embedding--based paradigms can conduct network construction and network inference via embedding highly diverse, heterogeneous, high-dimensional node information into a low-dimensional space. This is characterised with the highest level of complexity in terms of network representation and dynamics.

\textbf{Interpretability} represents the ability to explain or to present in understandable terms to a human \cite{IEEEexample:doshi2017towards}. The interpretability of networks is about the understanding of network representation and the corresponding network dynamics. It decreases as the network complexity increases with more information represented and modelled. From a local view, more complex rules of network formation enable smaller information loss but may result in less interpretable global emerging characteristic behaviours observed in networks. For example, compared with rule--based paradigms, agent--based paradigms may generate less interpretable networks as they incorporate the impact of various observables from microscopic perspective of agents \cite{IEEEexample:cazabet2011simulate}. From a global view, more complex models can embed more complex networked information but are characterised by less interpretable embedding process and inference results, such as the network embedding paradigms which are less interpretable in encoding or decoding but can represent highly diverse and heterogeneous networks.

\textbf{Measurability} represents the ability to measure a characteristic of a class of objects \cite{IEEEexample:rossi2007measurability}. Measurability of networks generated for abstract model's aims is about measuring the similarity between the inner rules of a networked system and that of real networks, while the measurability of networks under specific model's aims focuses on the measurable output for external tasks. Networks of higher interpretability but lower complexity, like non-attributed static networks generated by rule--based or graph--based paradigms, can be easily analysed, measured and compared in terms of network components and dynamics. These networks are usually involved in researches with abstract model's aims including topological feature analysis \cite{IEEEexample:barabasi1999emergence,IEEEexample:fortunato2006scale,IEEEexample:musial2013kind,IEEEexample:zhang2021vulnerability} and mimics of reality \cite{IEEEexample:euler1741solutio,IEEEexample:hunter2008goodness}. Networks of high complexity but low interpretability like the attributed temporal networks are always generated by event--based, agent--based, probabilistic graph--based or network embedding--based paradigms. They incorporate rich information especially in structural and temporal dimensions while preserve enough dynamics complexity for the fulfilment of specific model's aims including link prediction, node classification, community discovery, anomaly detection, synchronization and controllability.

\section{How to model the dynamics in networked systems}
\label{quadrant}
In this section, we explore the answers to the question "how?" about the modelling of dynamics in complex networked systems. This involves a discussion about two key concepts of complex networked systems: (i) dynamics of networks and (ii) dynamics over networks.
Dynamics of networks considers the network's generation process and network's changes over time (evolution), while the dynamics over networks refers to the dynamic processes that occur on networks over time. In addition to the process and network complexity, we also emphasise the complexity resulting from various interrelations between dynamic processes and network dynamics, where unilateral or mutual influence from one-way or two-way interactions are reviewed and discussed.
Given complex networks (obtained via methods from section \ref{how1}) and observable information of dynamic processes, their dynamics and interrelations should be modelled in a way that models' aims (see section \ref{aim}) are fulfilled.

\subsection{An overview of dynamics and their interrelations}
Taking the networks obtained via modelling paradigms introduced in section \ref{how1} as a starting point, the modelling of dynamics within the CNS, at the appropriate complexity level as per complexity dimensions introduced in section \ref{NT}, enables a deeper understanding of changeable network structures and processes over them with and contributes to achieving certain models' aims under the constraints of observability.

\begin{figure*}[t!]
	\centering
		\begin{minipage}[b]{0.45\linewidth}
			\includegraphics[width=1\linewidth]{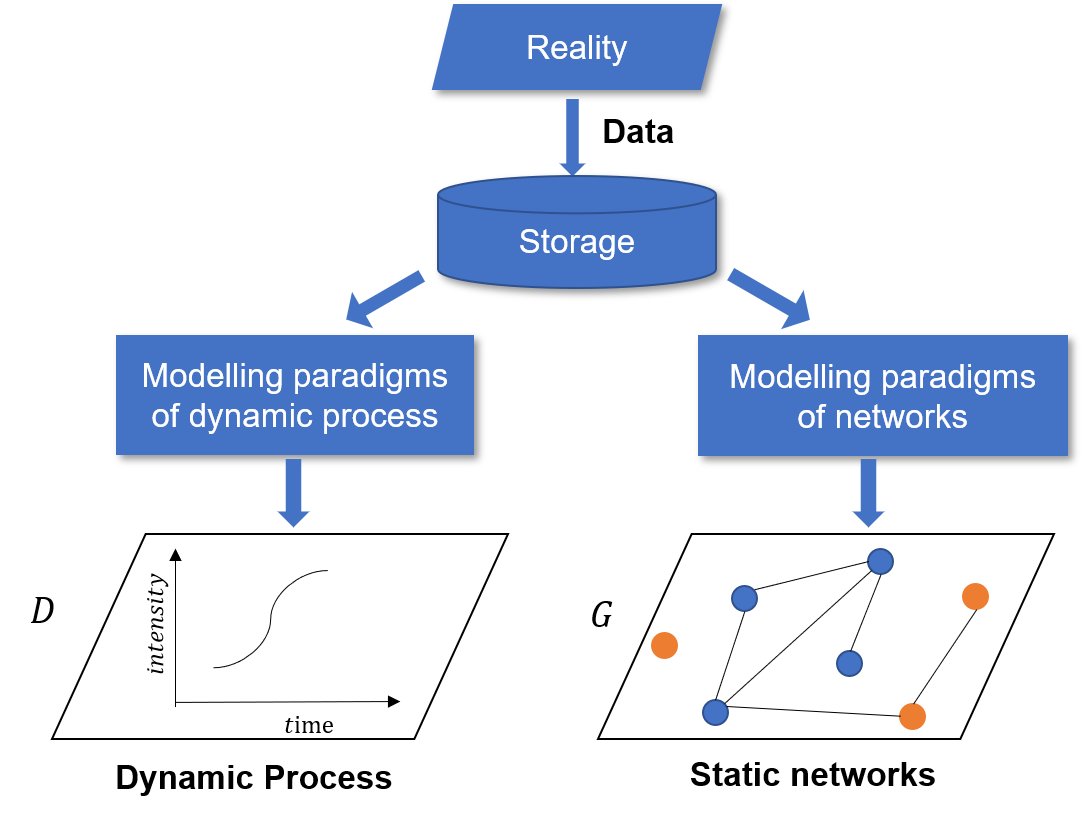}
			\subcaption{Generation 1}
	\end{minipage}\\
		\begin{minipage}[b]{0.45\linewidth}
			\includegraphics[width=1\linewidth]{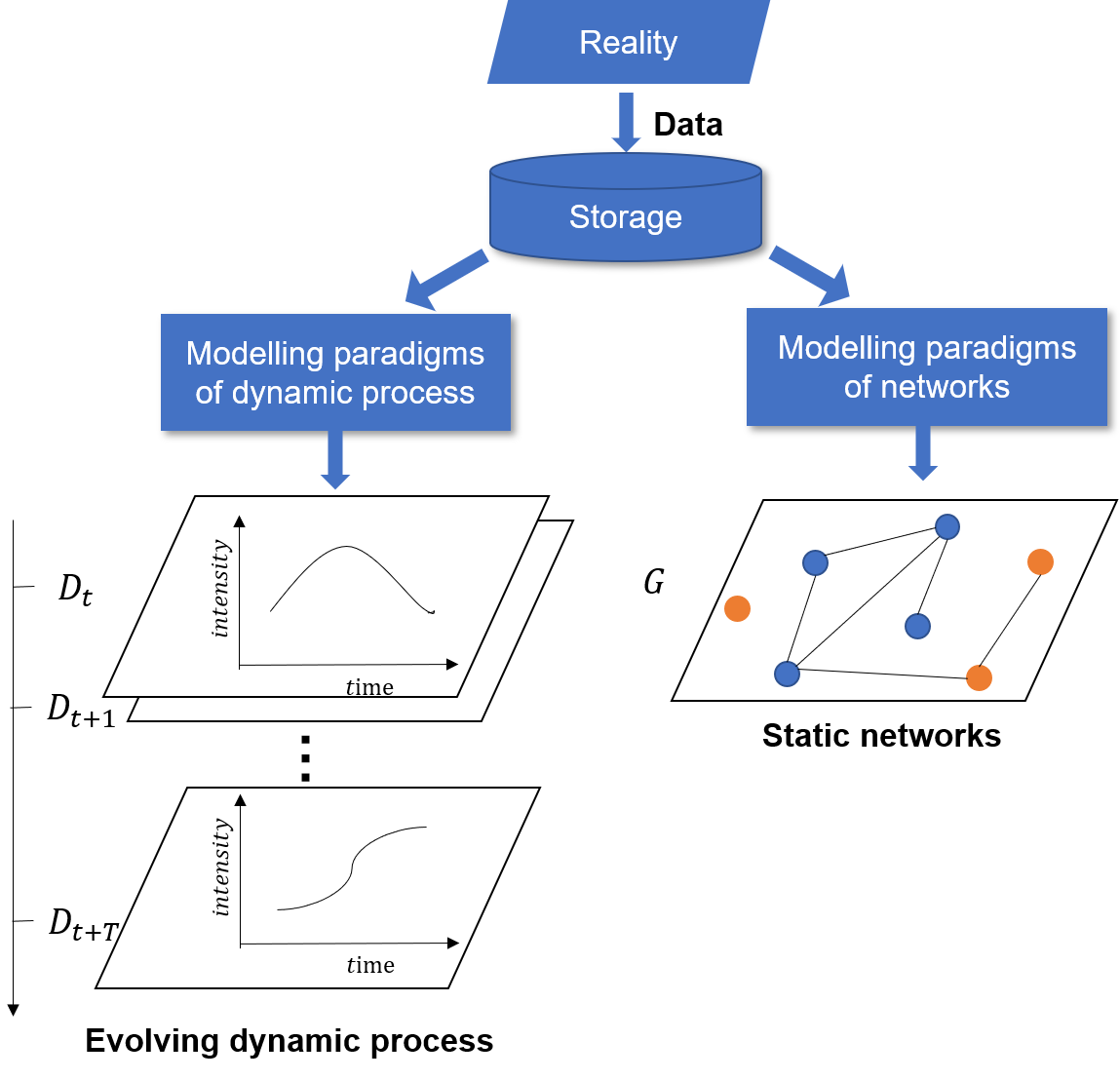}
			\subcaption{Generation 2a}
	\end{minipage}
		\begin{minipage}[b]{0.45\linewidth}
			\includegraphics[width=1\linewidth]{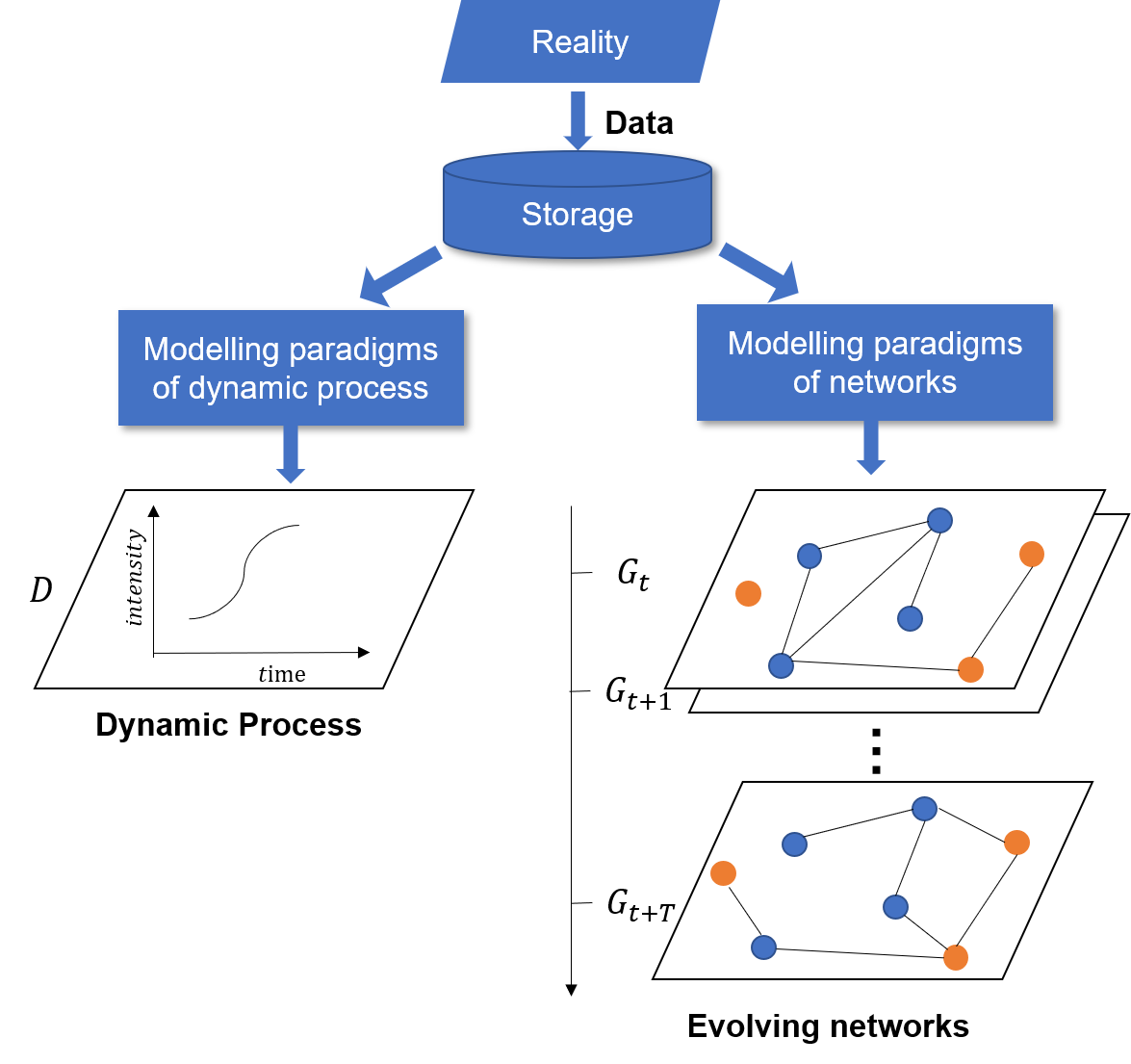}
			\subcaption{Generation 2b}
	\end{minipage}\\
		\begin{minipage}[b]{0.45\linewidth}
			\includegraphics[width=1\linewidth]{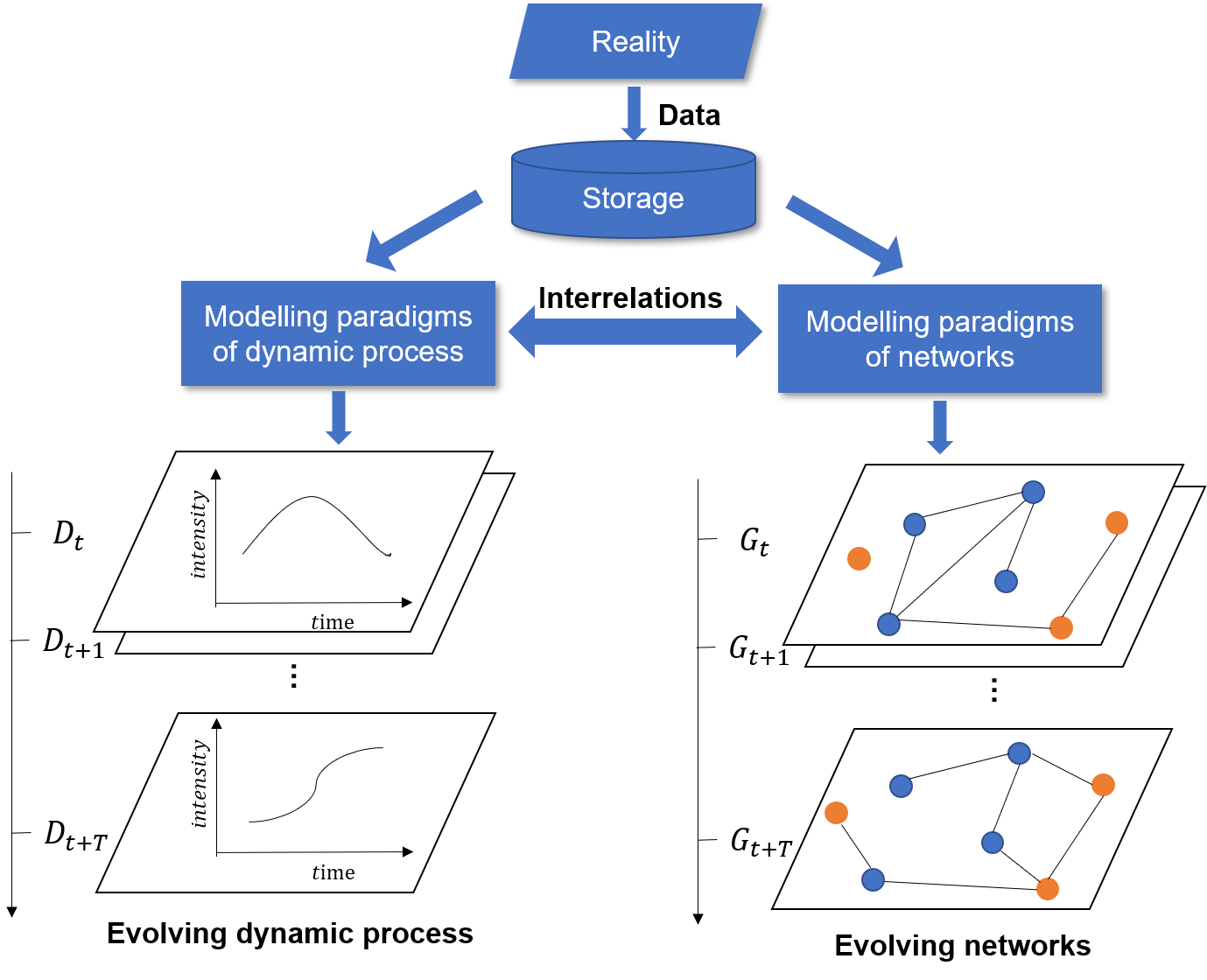}
			\subcaption{Generation 3}
	\end{minipage}
		\begin{minipage}[b]{0.45\linewidth}
			\includegraphics[width=1\linewidth]{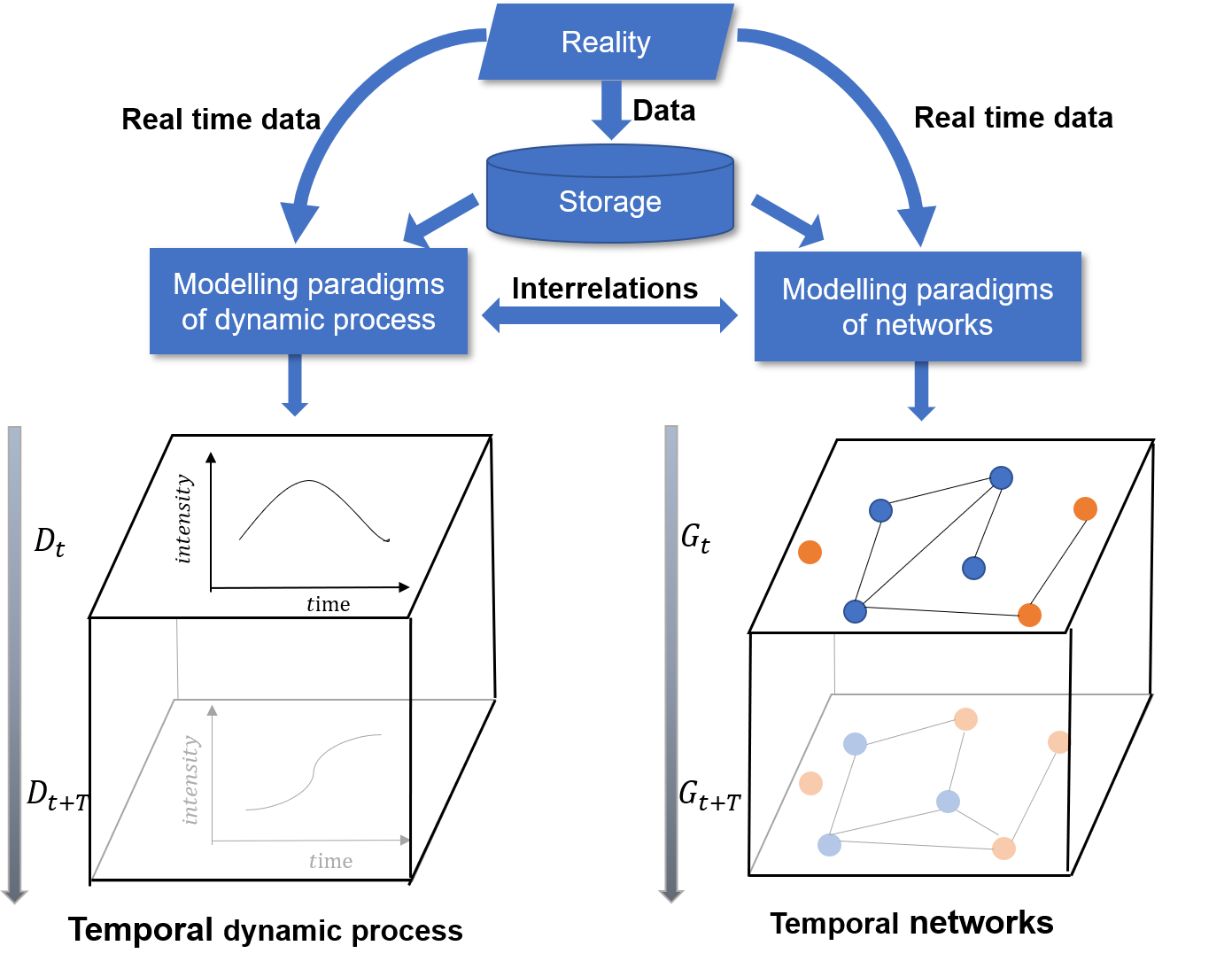}
			\subcaption{Generation 4}
	\end{minipage}\\
	\caption{Generations of modelling framework}
\label{gen}
\end{figure*}

The complexity of modelling CNS dynamics comes from three primary elements: the modelling of networks, processes over networks, as well as the interrelation between the two. To have an overview of various CNS resulting from those three elements, we introduce a 4-generation modelling framework (see Figure \ref{gen}) to navigate a pathway through different levels of complexity of CNS.


As is shown in Figure \ref{gen}a, in the first generation (generation 1), the research concentrates on static networks, where spreading processes are introduced without changing their parameters. The static networks with fixed nodes and edges just provide necessary spatial or structural information rather than one-way influence that might trigger the parameter changes (evolution) of dynamic processes. Some studies employ spatial networks that utilize a metric to embed spatial information \cite{IEEEexample:buhl2006topological,IEEEexample:barthelemy2011spatial}, where the dynamic processes are modelled under the constraints of space, like the human activity on transportation networks \cite{IEEEexample:buhl2006topological}. Majority of research on dynamic processes over complex networks can be classified as the first generation of models \cite{IEEEexample:brodka2020interacting}.

Assuming a virus $a$ spreading in the society where all relevant information is observable, in the generation 1, we can simplify this scenario into an epidemic spreading process of virus $a$ with fixed parameter (e.g. infection rate) on a static non-attributed networks that are just built with fixed nodes (e.g. people) and edges (e.g. social contact). This simplified CNS increases complexity in structural and spatial dimensions once we incorporate more information about node attributes (e.g. age, gender, location, etc.) and edge attributes (e.g. direction, weights, distance, etc.).

With the temporal complexity introduced in CNS in the second generation (generation 2), parameter changes of dynamic processes or network topology changes over time can be observed via snapshots in discrete time steps. In the generation 2a, CNS represents the interactions via evolving dynamic processes (captured in snapshots) and static networks (see Figure \ref{gen}b), where the spreading process on static networks can evolve through parameter changes \cite{IEEEexample:eletreby2020effects}. In the generation 2b, CNS represents the interactions directly via evolving networks (captured in network snapshots) and dynamic process with no parameter changes (see Figure \ref{gen}c). In this context, researchers aim to transform the snapshots of CNS into latent states and model their transition process discretely, involving the networks that switch arbitrarily between different adjacency matrices according to stochastic mechanisms like positive linear switched systems \cite{IEEEexample:rami2013stability} and Markov switching rules \cite{IEEEexample:sanatkar2015epidemic,IEEEexample:zhao2015dynamic}.

Taking the above mentioned spreading of virus $a$ as an example, in the generation 2a, a parameter (e.g. infection rate) of virus $a$ can change just arbitrarily or according to an external factor outside of CNS (e.g. temperature, etc.), while in the generation 2b, edge addition (e.g. social contact over time) or edge removal (e.g. enabled by the implemented policies of social distancing) can naturally lead to observations of evolving networks.

The modelling framework steps into the third generation(generation 3) as the dynamic process starts to co-evolve with the evolving networks (see Figure \ref{gen}d). The involved evolving dynamic processes are captured via evolving snapshots with discrete parameter changes, where the changes of parameters and the transition of the latent states of evolving networks can either be independent or interrelated. The interrelations, including unilateral and mutual influences, enable the modelling of co-evolution of dynamics and greatly increase the complexity of modelling CNS. For example, \cite{IEEEexample:gu2017co} model the co-evolution process of dynamic social networks and the opinion migration on networks via introducing mutual influence between these two dynamics, where opinion migrates based on social structure and social networks evolve considering the similarity of opinions. Generation 3 can easily collapse into the generation 2a when there is only one snapshot of dynamic processes and no interrelation is considered and vice versa for the generation 2b.

To have a better illustration of the interrelation in the generation 3, we also start with the simplest scenario where the virus $a$ spreads on a static social network (a single snapshot of evolving networks). In the generation 3, we can allow the network topology to change over time in response to the spreading of virus $a$ (e.g. people die of virus $a$ and are removed from the social networks). These evolving networks are interrelated with dynamic processes and are characterised with increasing temporal complexity. The parameter (e.g. infection rate) of virus $a$ can also change according to the network topology (e.g. number of social contacts, etc.) or node attributes (e.g. age, gender, etc.). In this way, the complexity of the CNS also increases with an introduction of the evolving dynamic process and interrelations. The CNS will become even more complex if we further incorporate a vaccination $b$ in the modelling framework of the generation 3, where the spreading of the vaccination $b$ can affect the node attribute (e.g. vaccinated or not) and directly change the parameter (e.g. infection rate) of the virus $a$.

The modelling framework then goes through a fundamental change in the fourth generation (generation 4) as the time gaps between the CNS snapshots are narrowed, in the limit, to zero and the co-evolving dynamics are represented and modelled continuously with the introduction of real time data. As is shown in Figure \ref{gen}e, based on both historical data from a data storage and the real time data, the temporal networks capture and model all the instantaneous interactions, while the temporal dynamic processes are modelled with continuous parameter changes. The complexity of the CNS increases as more complex interrelations between continuous dynamics are introduced and modelled. There are studies on modelling continuous changes of networks via introducing ordinary differential equations (ODE) to the GNN methods \cite{IEEEexample:huang2021coupled}, but none of them involve the most complex case of modelling continuous co-evolving dynamics of interrelated dynamic processes and dynamic networks. CNS in the generation 4 can approach the ultimate goal of DTs when the model output of CNS is additionally fed back to and can influence the reality in a real time manner as a reference for practice. We refer to this scenario as the fifth generation of modelling framework (generation 5), where a closed feedback loop of real time monitoring, simulation, forecasting and deriving solutions for reality is formed and enables the CNS to approach DTs as an extension of reality.

To be more specific and continuing with our illustrative example, in the generation 4, the constant event streams about the spreading of the virus $a$ and the temporal networks can be observed and modelled simultaneously. The instantaneous social contact and the virus $a$ spreading can be captured instantly. CNS in this context can react to sudden changes observed in reality and evolve in real time. Any changes to the temperature, recorded all the time, can trigger continuous parameter (e.g. infection rate) changes of the spreading virus $a$. The vaccination $b$ can also be introduced at any time and trigger network attribute (e.g. vaccinated or not) change right away. In generation 5, the simulation result of CNS about the spread of the virus $a$ can be fed back to the policy makers and trigger, for instance, a launch of a "promote vaccination $b$" campaign, where the spread of vaccination $b$ will be simultaneously monitored and modelled by CNS with a real time output about social networks fed back to the policy makers.

The above mentioned modelling framework shows with examples of real scenarios that CNSs can be represented and modelled with increasing complexity through generations and finally reach the goal of a DT in generation 5. To be more aware of the progress of studies on CNSs under this modelling framework, we need to review and discuss what kind of model's aims can be achieved by modelling the dynamics in networked systems, and how these dynamics are modelled in network dimension, process dimension and both of these dimensions.

\subsection{Modelling Dynamics of Networks}
Dynamics of networks can result in networks with different structural characteristics and their changes over time. Their modelling fulfils models' aims via analysing and learning these resulting patterns and changes of nodes, edges and attributes. There are overlapping areas between modelling dynamics of networks (in a way that models' aims can be fulfilled) and modelling paradigms of network generation (in a way that minimises information loss), as a good model that approaches real network dynamics can naturally accomplish these two tasks simultaneously. Studies on modelling dynamics of networks start from patterns of non-attributed static networks and then turn to dynamic networks that consist of nodes and edges that change over time. Those networks include temporal information about characteristics and behaviours of their components. Time dimension is introduced to break the assumption of static networks with fixed nodes and edges \cite{IEEEexample:skarding2020foundations,IEEEexample:ranshous2015anomaly,IEEEexample:rossetti2018community}.

Modelling of the network dynamics, considering the patterns and changes of nodes, edges or attributes from a structural view, can be categorised as models that allow for: (i) no changes, (ii) topology changes, (iii) attribute changes, (iv) both attribute and topology changes and (v) structural pattern changes. They differ in whether network structures change and how they change over time, each to varying degrees coping with the models' aims including prediction and classification of network components, as well as the pattern discovery of network structures. As networks with only attribute changes are only involved in studies of dynamic processes on networks, we only discuss models of network dynamics for category (i), (ii),(iv) and (v) just from the perspective of a network dimension.

\subsubsection{No changes}
\label{nochange}
The modelling of network dynamics starts from static networks without any changes of structural components over time and focuses on the prediction of unobservable network components.

Networks built only with fixed nodes and edges are of the least complexity within this no change category, which involves missing link prediction fulfilled with approaches that are applicable for static networks.
Some techniques take a link prediction of these networks as an unsupervised ranking problem based on a score for each non-observed link, which can be calculated either with a structural similarity index or probabilistic and statistical functions \cite{IEEEexample:martinez2016survey}. Structural similarity--based methods assume that nodes tend to form links with other similar nodes, which solely consider the local or global topological information of networks \cite{IEEEexample:lu2011link}. Probabilistic and statistical function--based models abstract the network structure and then predict probabilities of the missing links using the learned model \cite{IEEEexample:martinez2016survey,IEEEexample:wang2019community}, where rule--based modelling paradigms in section \ref{how1.1} and probabilistic graph--based paradigms mentioned in section \ref{how1.2} can be employed to fit network topology for a probability score of each link.
There are also techniques that take link prediction as a supervised binary classification problem about whether each pair of nodes are connected or not, where machine learning models like logistic regression and decision tree can be applied to account for the effects of different topological similarity metrics \cite{IEEEexample:al2011survey,IEEEexample:bu2019link}. Another universally applied supervised approach is the network embedding modelling paradigm mentioned in section \ref{how1.2}, where the low-dimensional latent space representation of nodes is learned and their connections can be inferred via dependencies of latent space.

Networks increase in structural complexity when attributes are gradually introduced for nodes and edges.
Some approaches use node attributes to assist prediction of links \cite{IEEEexample:lu2009similarity}, where above introduced basic approaches for link prediction have been altered to incorporate node features. For unsupervised ranking algorithms, as is summarised by \cite{IEEEexample:al2011survey}, the similarity score between two attributed nodes can be calculated to assist a link prediction using methods including vertex feature aggregation \cite{IEEEexample:barabasi2002evolution}, kernel feature conjunction \cite{IEEEexample:basilico2004unifying}, extended graph formulation \cite{IEEEexample:al2006link} and generic SimRank method \cite{IEEEexample:jeh2002simrank}. In addition, probabilistic and statistical function--based models can also incorporate attributes to model the probability of links between each pair of nodes \cite{IEEEexample:al2011survey}. Supervised models that take link prediction as a binary classification problem can also be improved based on the above mentioned information about attributed nodes. For example, \cite{IEEEexample:al2006link} introduces node attributes via vertex feature aggregation to the machine learning algorithms like a decision tree or SVM in the link prediction tasks. There are also improved network embedding methods extensively reviewed by \cite{IEEEexample:cui2018survey} for the link prediction task considering an effect of both network topology and node attribute. An example of such an approach is the deep attributed network embedding method designed by \cite{IEEEexample:hong2019deep} using a deep neural network based on the topology proximity and attribute proximity.

Some studies additionally introduce attributes to the edges and transform the link prediction into a classification task for multi-relational networks. Once the transformation is completed, this problem can be approached by the above mentioned probabilistic and statistical function--based approaches and supervised learning approaches considering the similarity of nodes and dependencies of inner principles. For example, \cite{IEEEexample:taskar2003link} uses relational Markov network to investigate the probability of link labels given the known node attributes. \cite{IEEEexample:dai2017link} performs link prediction in multi-relational networks using a non-negative matrix factorization algorithm based on relational similarity.

Other methods introduce attributes to the networks and focus on the node classification tasks, which employ models considering the effect including node attributes \cite{IEEEexample:gao2018deep,IEEEexample:hong2019deep,IEEEexample:liu2021block} and edge attributes such as weights \cite{IEEEexample:taskar2001probabilistic,IEEEexample:kleinberg2002approximation,IEEEexample:hong2019deep,IEEEexample:mcsherry2001spectral,IEEEexample:hong2019deep}. Given the sparsity of graphs with fully labelled nodes and the time consuming manual labelling, most studies employ partially labelled graphs and train a classifier for the prediction of unlabelled nodes.
Referring to the models for a node classification, very well summarised by \cite{IEEEexample:bhagat2011node}, we further categorise these modelling approaches into three types: (i) unsupervised learning approaches including probabilistic and statistical relational learning \cite{IEEEexample:taskar2001probabilistic}, metric modelling \cite{IEEEexample:kleinberg2002approximation}, spectral partitioning \cite{IEEEexample:mcsherry2001spectral} and graph clustering \cite{IEEEexample:zhou2009graph}; (ii) supervised learning approaches that build classifiers like logistic regression model given features of labelled nodes \cite{IEEEexample:gao2018deep}; and (iii) the semi-supervised learning approaches given the labels of few nodes, where network embedding methods like random walk--based \cite{IEEEexample:zhu2003semi} and GCN--based Network embedding \cite{IEEEexample:hong2019deep,IEEEexample:abu2020n} are employed to account for both topological and attribute proximity effect \cite{IEEEexample:hong2019deep}.

\subsubsection{Topology changes}
\label{topologychange}
The modelling of network dynamics becomes more complex as network components start to change over time. The changes of topology include the edge addition and removal, node addition and removal. As the network formation and evolution can be captured in a time series of static snapshots, modelling approaches of link prediction for static networks mentioned in section \ref{nochange} can be extended and improved to learn one or several types of topology changes.

Studies on network topology changes start from networks built with fixed nodes and addition of edges, where the edges can be predicted and inferred via edge formation process extracted from networks over time using structural similarity--based models or probablistic and statistic function--based models discussed in section \ref{nochange}. For example, \cite{IEEEexample:wang2007local} employ a Markov model based on both topological and semantic features similarity between two nodes to evaluate the probability of a link. \cite{IEEEexample:wahid2019predict} predict formation of new links based on a combined popularity and similarity measure, which incorporates both global and local topological information via the introduction of Newton’s gravitational law.

Modelling network dynamics becomes more complex when it comes to networks built with fixed nodes and temporal edges that are added or removed over time. Link prediction tasks for these networks focus on the temporal topology information and its evolution, where modelling approaches for prediction of missing links in static networks have been improved to account for the effect of historical topology. For unsupervised similarity ranking--based models, the link prediction task is based on the predicted similarity scores calculated using past structural similarity scores via a time series forecasting model such as ARIMA \cite{IEEEexample:gunecs2016link}. To deal with the model capacity and computational efficiency problem for probabilistic and statistical function--based models, efficient learning algorithms can be introduced to account for influence of topology, such as the proposed neighbour influence clustering algorithm in a conditional temporal restricted Boltzmann Machine for a prediction of temporal edges \cite{IEEEexample:li2014deep}. As for supervised learning approaches, graph convolution network (GCN) is widely used to learn a node structure of a network snapshot for each time slide and LSTM is employed to performs a temporal feature learning for all the network snapshots \cite{IEEEexample:chen2018gc,IEEEexample:lei2019gcn}.
The discussed so far approaches only use network topology, but the heterogeneous prior information, such as node attributes, is suggested to be used to further improve the accuracy \cite{IEEEexample:ma2018graph} but is not commonly explored in the research community. \cite{IEEEexample:sarkar2014nonparametric} propose a nonparametric link prediction algorithm that can use both topology and node labels for the calculation of linkage probability with seasonality linkage patterns.

There is an even more complex case when the networks shrink or grow in size as they evolve with temporal nodes and edges. However, as the networks can be modified to include all observed nodes from all the snapshots of temporal networks, it is always assumed that there is a fixed set of nodes for all the networks at different time points \cite{IEEEexample:chen2018exploiting}, where the above mentioned modelling approaches for link prediction of networks built with fixed nodes and temporal edges can be used.

To conclude, the existing studies focus on modelling topology changes where nodes are fixed. The current methods consider one of the two scenarios: (i) edge addition and (ii) edge addition or removal. There is space for further research on modelling networks with temporal nodes and edges, where addition and removal of nodes, as well as the resulting change of network sizes should be considered.

\subsubsection{Attribute and topology changes}
The modelling of network dynamics becomes more complex as we allow, on top of the topology changes, for network attributes to change over time. As current models of topology evolution are limited to networks built with a fixed set of nodes and edge changes over time, the models of networks where both attributes and topology change also have the same limitation. These models, given an addition or removal of edges, consider: (i) edge attributes changes, (ii) node attributes changes and (iii) both of these changes. Modelling approaches for the link prediction and node classification of static networks presented in section \ref{nochange} and the network topology changes introduced in section \ref{topologychange} can be extended to learn both attribute changes and topology changes.

The topology change, linked to the edge addition or removal, can be accompanied with changes of edge attributes. There are studies on link prediction in temporal networks that have edge weights \cite{IEEEexample:chen2018exploiting,IEEEexample:ma2018graph,IEEEexample:lei2019gcn} and directions \cite{IEEEexample:chen2018exploiting}. The typical supervised learning approach for topology changes mentioned in the section \ref{topologychange}, where GCN explores the local topology of each snapshot and LSTM characterises the evolving features of dynamic networks, can be improved by introducing the generative adversarial network (GAN) to tackle the sparsity and the wide-value-range problem of edge weights \cite{IEEEexample:lei2019gcn}. Network embedding methods based on matrix factorisation can also be improved to include the information about edge weight or direction into adjacency matrix for a prediction of temporal edges \cite{IEEEexample:ma2018graph,IEEEexample:chen2018exploiting}. Further research is needed for a variation of other edge attribute changes.

When node attributes change in networks built with fixed nodes and non-attributed edges that change over time, modelling approaches of node classification can be employed to learn the changes of the node attributes by considering the evolution of attributes and topology.
\cite{IEEEexample:ashraf2019simulation} use GCN to conduct node classification task on the social networks built with a fixed number of attributed nodes, changeable node labels and non-attributed edges that are added over time. GCN implemented in this research not only considers the local topology and its attributes, but also uses the similarity--based matrices to account for patterns of high-order neighborhood. Further research is needed for node classification in networks built with dynamic labelled nodes and temporal edges.
There also comes the most complex case where node attributes, edge attributes and network topology all can change at the same time. Currently, there is no research in this space and further studies are required to model this very complex scenarios.

\subsubsection{Structural pattern changes}
The above mentioned structural components and their changes result in various structural patterns and corresponding changes. Structural patterns refer to the correlated combination of nodes, edges and attributes within a community or a network. Research in the space of structural patterns includes patterns and their dynamics discovery, analysis and prediction. The common models' aims here are e.g. a community discovery and an anomaly detection.

A community discovery starts from defining a community. This characterises the structural patterns of the sub-networks to be discovered via generally unsupervised way of modelling. A community in a complex network, as is defined by \cite{IEEEexample:coscia2011classification,IEEEexample:rossetti2018community} in a generic way, is a set of entities that share some closely correlated sets of actions with the other entities of the community. To fulfil the requirement of reflecting certain features of reality, the closeness within each community can be measured based on density, vertex similarity, actions of nodes or influence spread, all corresponding to different types of community discovery algorithms well summarised by \cite{IEEEexample:coscia2011classification}.

Considering the increasing temporal and structural complexity of networks resulting from the topology variations and an introduction of attributes, the community discovery approaches have also been further developed and categorised from the perspective of process that is seen as the inner structures of algorithms. To deal with the community instability problem in dynamic networks, the temporal smoothing operations have been included in community discovery models to smooth-out the evolution of communities, which accordingly involve a new category based on varying extent of temporal-smoothness \cite{IEEEexample:rossetti2018community}. For attributed networks, built with attributed nodes and edges, a fusion procedure has been introduced to community discovery models to account for both effects of topology and node features, where another category based on when and how they use and fuse network structure and attributes is proposed \cite{IEEEexample:chunaev2020community}. To deal with the directed networks that are featured with edge directions and asymmetrical matrices, different community discovery models have been proposed and well summarised depending on the way directed edges are treated \cite{IEEEexample:malliaros2013clustering}.

However, the perfect community discovery algorithm does not exist despite of all the above mentioned attempts, as each of them performs well on one specific declination of the general problem and can achieve different partitions even for the same networks \cite{IEEEexample:rossetti2018community}. Based on that, \cite{IEEEexample:coscia2019discovering} further categorise the community detection algorithms according to the similarity of their results, which  attempts to confirm the valid definitions of a community and help with the choice of algorithms for future research. There is still a need for further research on the community discovery of network variations with varying degrees of temporal complexity and structural complexity, given the superposed challenges of community definition, temporal smoothness as well as topology and attribute information incorporation.

Studies on anomaly detection focus on the rare occurrences of structural components, patterns as well as their changes, involving detection of anomalous nodes, edges, subgraphs, events and graphs \cite{IEEEexample:ranshous2015anomaly,IEEEexample:kendrick2018change,IEEEexample:ma2021comprehensive}. They start from static networks built with fixed nodes and edges, where anomaly detection of nodes, edges and subgraphs can be realised via traditional non-deep learning approaches based on the network statistical features or using representation learning methods, as is summarised by \cite{IEEEexample:akoglu2015graph,IEEEexample:ma2021comprehensive}. These methods have also been improved respectively when it comes to attributed networks with richer information about network structures \cite{IEEEexample:akoglu2015graph,IEEEexample:ma2021comprehensive}.

As temporal complexity is introduced to networks, two types of anomaly detection methods can be distinguished. More specifically, there is a two-stage approach that maps networks into a vector of real numbers and then employs an anomaly detector on it for a node, edge or subgraph anomalies, involving well categorised community, compression, decomposition, distance, and probabilistic model--based models \cite{IEEEexample:ranshous2015anomaly}. There are also deep learning approaches respectively applied to the anomaly detection of nodes, edges, subgraphs and graphs \cite{IEEEexample:ma2021comprehensive}. There is also research gap that calls for further study on anomaly detection models incorporating both attribute and temporal information of networks.

\subsection{Modelling Dynamic Processes}
In this subsection, we focus on the dynamic processes that can take place over networks. These dynamic processes interact with the dynamics of networks and vice versa. Dynamic processes can result in three types of changes: (i) a network topology change; (ii) a network attribute change; and (iii) a parameter change of dynamic processes \cite{IEEEexample:funk2010modelling}. The network topology change and the attribute change can also lead to the parameter change of dynamic processes \cite{IEEEexample:sun2021competition}. There are many interesting variations of the combination of dynamic processes and dynamic networks, which differ depending on the research objectives and application scenarios. Corresponding studies start from single parameterized dynamic processes on the static networks and then turn to multiple dynamic processes with dynamically changing parameters on the dynamic networks, which is featured with increasing process complexity, network complexity, as well as the complexity resulting from various interrelations between dynamic processes and network dynamics

Modelling of the dynamic processes research can be categorised into the following groups: (i) a single dynamic process, (ii) independent multiple processes and (iii) interrelated multiple processes. They differ in a number of dynamic processes and how they interact, each to a varying degree mimicking the real world. Within each category, researchers either focus on parameterized dynamics or dynamically changing dynamics.

\subsubsection{Single dynamic processes}
\label{single}

\textbf{Spreading dynamics} on complex networked structures obtained via methods in section \ref{how1}, as a typical dynamic process, is covered by considerable literature on the spread of phenomena/medium ranging from virus \cite{IEEEexample:ameen2020efficient,IEEEexample:sahneh2014competitive,IEEEexample:cooper2020sir,IEEEexample:syafruddin2013sir}, meme \cite{IEEEexample:ye2018open}, opinion \cite{IEEEexample:amblard2004role,IEEEexample:cho2012identification,IEEEexample:klamser2017zealotry}, decision making \cite{IEEEexample:xu2018exploiting}, nutrition \cite{IEEEexample:laur2018sustain,IEEEexample:kendig2020host} and social contagion \cite{IEEEexample:karampourniotis2015impact}. 


The modelling of spreading dynamics on networks starts from classic population models with the simplest analysis that considers the evolution of the state for all individuals rather than the state of each individual \cite{IEEEexample:nowzari2016analysis}. They include the stochastic population model that describes the evolution of the population state via a Markov process, as well as its approximation, deterministic population model with deterministic definitions of the population state \cite{IEEEexample:nowzari2016analysis,IEEEexample:zino2021analysis}. To model the states of each individual independently and allow for arbitrary interactions among them, as is summarised by \cite{IEEEexample:nowzari2016analysis}, these population models are extended and improved to faithfully learn spreading processes on networks.

A spreading process on static networks is an example of a simplified representation of various real world scenarios where networks underlying the dynamic processes are simulated or represented under the most stringent assumption of nodes and edges that do not change. When a dynamic process evolves much faster than the network of interactions, static networks can serve as accurate proxies of slowly switching topologies. This real world situation can be approximated as dynamic processes over static networks \cite{IEEEexample:zino2021analysis}.

The modelling of dynamic processes on static networks start from analysing the impact of pairwise interactions using the extended classic models, including stochastic network model and deterministic network model \cite{IEEEexample:nowzari2016analysis,IEEEexample:zino2021analysis}, where networks with varying features and structures can be introduced to the studies.
Stochastic network models assume the state transition for each node as a Markov jump process or its extension under relaxed Markovian assumptions \cite{IEEEexample:kiss2015generalization,IEEEexample:liu2018burst,IEEEexample:melo2018knowledge,IEEEexample:zino2021analysis}. Deterministic network models, as approximations of stochastic network models with deterministic definitions of node states, are also widely used \cite{IEEEexample:van2011n,IEEEexample:ahn2013global}. A threshold model is one example of a typical deterministic model for an epidemic process \cite{IEEEexample:karampourniotis2015impact}. The modelling of spreading dynamics becomes more complex and realistic as complex contagion is also considered with the incorporation of group interactions, which currently has been realised via simplicial complexes \cite{IEEEexample:iacopini2019simplicial}.

A spreading process on dynamic networks refer to one of the most common real-world application scenarios. The underlying networks of interactions are not static, but dynamically change while co-evolving with and influenced by the dynamic processes over the networks \cite{IEEEexample:gross2008adaptive}. 
Models that capture their co-evolution at comparable time-scales have been well categorised into temporal-switching, activity-driven,
and edge-Markovian networks \cite{IEEEexample:zino2021analysis}. For dynamic processes on dynamic networks, temporal switching networks model them as snapshots switching arbitrarily between a set of topologies according to stochastic mechanisms such as Markov switching rules \cite{IEEEexample:sanatkar2015epidemic,IEEEexample:zhao2015dynamic}.  Activity-driven approaches focus on the networks' interactions generated according to a time invariant function characterising individual properties, which involves a series of extensions with the introduction of epidemic threshold due to its analytical tractability, as is summarised in detail by \cite{IEEEexample:leitch2019toward}. Edge-Markovian dynamic graphs can model stochastic evolution of dynamic networks, which also involves analytically tractable extensions with spreading dynamics \cite{IEEEexample:kiss2012modelling,IEEEexample:taylor2012epidemic}.

Spreading dynamics on dynamic networks can also be modelled with data-driven machine learning approaches. They focus on transforming spatial information and other temporal features involved in a spreading process to well handled temporal information, where deep learning--based predictive models like Recursive Neural Networks and Convolutional Neural Networks can be employed to predict spreading process, as is well summarised by \cite{IEEEexample:baldo2021deep}. Researchers also start to employ network embedding approaches, as is mentioned in section \ref{how1.2}, to incorporate network information into the predictive systems, which involves typical examples of predicting epidemic spreading with graph neural networks \cite{IEEEexample:mevznar2020prediction} or using node regression based on transfer learning \cite{IEEEexample:mevznar2021transfer}.

\textbf{Parameter changes} of processes are discussed in terms of the above mentioned single dynamics. The existing modelling approaches to a single dynamic process propagating over the network are generally parameterized without any change of parameters implemented. Only few studies focus on the evolution of spreading processes on networks. \cite{IEEEexample:eletreby2020effects} incorporate mutation of pathogen strains and corresponding changes of epidemic transmission probability, which trigger the evolutionary adaptations of the spreading processes with dynamically changing epidemic threshold. For this example, transmissibility changes between limited number of fixed values and is controlled via mutation and transition probabilities. 

\subsubsection{Independent multiple processes}
\label{independent}
Dynamic processes can result in three types of changes, including: (i) a network topology change; (ii) a network attribute change; and (iii) a parameter change of dynamic processes \cite{IEEEexample:funk2010modelling}. As almost all the models of multiple processes 
in section \ref{single} are based on probability of state transition, the parameters of multiple processes in section \ref{independent} and section \ref{interrelated} refer to the transmissibility, adoptability or probability of an entity being infected or activated or an entity adopting a given behaviour or state.

Independent multiple processes take place independently without direct influence on their parameters as they ignore the dependence of multi-spread in co-infected status or just exclude the concurrent infections of multi-spread. Since no research can be found on independent multiple processes where co-infected status is allowed, we mainly focus on independent multiple processes that exclude co-infected status and interact via changes of network topology, attributes or structural patterns.

\textbf{Mutually exclusive processes} refer to the multiple processes propagating over the network where the concurrent infection by more than one spread is not possible. The goal is to investigate circumstances under which the dominance of a single spread can emerge\cite{IEEEexample:karrer2011competing,IEEEexample:newman2005threshold,IEEEexample:budak2011limiting}.

These dynamic processes interact on the same network while preserving their independence via assumptions including temporal separation \cite{IEEEexample:newman2005threshold}, structural separation \cite{IEEEexample:sun2021competition}, cross immunity \cite{IEEEexample:karrer2011competing,IEEEexample:prakash2012winner,IEEEexample:min2018competing} and cross adoption \cite{IEEEexample:fu2019analysis,IEEEexample:koprulu2019battle}. Under temporal separation, two pathogens can spread independently in separate time steps and interact via network topology changes, such as node removal as a result of death or immunity \cite{IEEEexample:newman2005threshold}. Studies on concurrent multiple processes generally adopt cross immunity or cross adoption assumption to deal with the concurrent infection that is not allowed for mutually exclusive competing processes. Under cross immunity, the states of an infected/recovered network vertex are changed to be immune to any other infections \cite{IEEEexample:karrer2011competing,IEEEexample:prakash2012winner,IEEEexample:min2018competing}. Under cross adoption, the infected network nodes can transit to be just infected by another spread with a specified probability \cite{IEEEexample:fu2019analysis,IEEEexample:koprulu2019battle,IEEEexample:pan2018effective}. There is also research on concurrent multiple processes that use structural separation, where network nodes are grouped and are only available for specific spreading dynamics, such as the simple contagion and the complex contagion that work out for predetermined vertex groups \cite{IEEEexample:sun2021competition}.

Under the above mentioned assumptions and basic settings, 
the models widely employed in studies of single spreading dynamics in section \ref{single} can be extended and used to model independent, multiple processes. In addition, there are already studies using $SI$ \cite{IEEEexample:sun2021competition}, $SIS$ \cite{IEEEexample:prakash2012winner}, $SIR$ \cite{IEEEexample:min2018competing,IEEEexample:fu2019analysis}, independent cascade model \cite{IEEEexample:budak2011limiting}, percolation model \cite{IEEEexample:newman2005threshold,IEEEexample:min2018competing} and Reed–Frost model \cite{IEEEexample:karrer2011competing}. For example, \cite{IEEEexample:sun2021competition} propose dynamic message-passing equations for two $SI$-type competing processes to incorporate the message-passing into the parameters. \cite{IEEEexample:prakash2012winner} use the extended model, $SI_1I_2S$, to model the propagation of two concurrent epidemic spreading processes.

\textbf{Parameter changes} of dynamic processes are discussed in terms of independent multiple processes. There are scenarios (e.g. election information spread and entertainment information spread in the same social networks) concerning parameterized independent multiple processes where co-infected status is allowed under the ignorance of the dependence of multi-spread without changing their parameters, though no research can be found here.

Most of the mutually exclusive processes involved in the existing studies do not consider dynamically changing parameters of the process \cite{IEEEexample:newman2005threshold,IEEEexample:karrer2011competing,IEEEexample:prakash2012winner,IEEEexample:min2018competing}. A small number of studies introduce dynamics with parameter changes over time under the impact of node states, where transmissibility varies from nodes' groups (structural patterns) \cite{IEEEexample:koprulu2019battle,IEEEexample:fu2019analysis}, states of neighboring nodes \cite{IEEEexample:pan2018effective}, message-passing \cite{IEEEexample:sun2021competition}. The parameters can either change between limited number of fixed values \cite{IEEEexample:koprulu2019battle,IEEEexample:fu2019analysis} or change continuously according to the network attributes resulting from another spreading process \cite{IEEEexample:sun2021competition}.

\subsubsection{Interrelated multiple processes}
\label{interrelated}
Interrelated multiple processes are characterised by direct unilateral or mutual influence of processes on their parameters. They can interact not only by changing network topology or attributes, but also via parameter changes.

\textbf{Partially inclusive processes} refer to the  multiple processes that allow the concurrent infections of nodes while also incorporate the dependence between spreads themselves. 

Interrelated multiple processes can have suppressing \cite{IEEEexample:funk2009spread,IEEEexample:pei2013spreading} or supporting \cite{IEEEexample:sun2021competition} relations, which are involved in only limited number of studies and so far all of them consider static networks \cite{IEEEexample:funk2009spread,IEEEexample:pei2013spreading,IEEEexample:sun2021competition}. These spreading dynamics change parameters with the transition of concurrent infection states. For example, epidemic spreading process, under the suppressing impact of awareness spreading, has different infection probabilities given nodes' different levels of awareness \cite{IEEEexample:funk2009spread,IEEEexample:pei2013spreading,IEEEexample:zhan2018coupling}. Similarly, collaborative multiple processes also have different probabilities under their supporting impact \cite{IEEEexample:sun2021competition}.

Under the above mentioned assumptions of concurrent infection, 
the models mentioned in section \ref{single} and section \ref{independent} can also be extended and used, where some of them have already been used to model interrelated multiple processes, including $SI$ \cite{IEEEexample:sun2021competition}, $SIS$ \cite{IEEEexample:zhan2018coupling}, $SIR$ \cite{IEEEexample:funk2009spread} and $SIS-SIRS$ \cite{IEEEexample:pei2013spreading}.

\textbf{Parameter changes} of processes here are discussed in the context of interrelated multiple processes. For almost all the existing research, at least one of the considered spreading dynamics has dynamically changing parameters in response to the impact of another spreading dynamics \cite{IEEEexample:funk2009spread,IEEEexample:pei2013spreading,IEEEexample:zhan2018coupling}. \cite{IEEEexample:sun2021competition} further introduce collaborative multiple processes that all change parameters with the message passing of nodes. 
Similarly to the case of independent multiple processes, parameters of interrelated multiple processes can either change between limited number of fixed values \cite{IEEEexample:funk2009spread,IEEEexample:pei2013spreading,IEEEexample:zhan2018coupling} or change continuously according to the network attributes resulting from another spreading process \cite{IEEEexample:sun2021competition}.

\subsection{Combination of the network and process dimensions}
In this section, we focus on the superposition of the network dimension and the process dimension, as well as the increasing complexity of modelling CNS considering the interactions and interrelations between the network and a dynamic process.

\subsubsection{Superposition of networks and processes}

Propagation process dynamics, either on static networks or dynamic networks, has been extensively studied using 
non-machine learning approaches \cite{IEEEexample:nowzari2016analysis,IEEEexample:zino2021analysis}, where 
data-driven machine learning approaches have recently been a popular choice of incorporating more structurally and temporally complex network information \cite{IEEEexample:mevznar2020prediction,IEEEexample:mevznar2021transfer}. Dynamic networks involved in these studies only allow for topology changes \cite{IEEEexample:sanatkar2015epidemic,IEEEexample:zhao2015dynamic,IEEEexample:mevznar2020prediction,IEEEexample:mevznar2021transfer} which influences the result of spreading dynamics.

The majority of the independent multiple processes considered in the existing and current research take place on static networks \cite{IEEEexample:sun2021competition,IEEEexample:karrer2011competing,IEEEexample:prakash2012winner,IEEEexample:min2018competing,IEEEexample:fu2019analysis,IEEEexample:budak2011limiting} and only a small number of studies can be found for those on dynamic networks \cite{IEEEexample:newman2005threshold,IEEEexample:koprulu2019battle}. As a typical example of competing epidemic processes, \cite{IEEEexample:newman2005threshold} allows removal of nodes and their edges over time as a representation of death or immunity and this results in the topology changes. However, interrelated multiple processes are involved in only a limited number of studies and so far all of them take place on static networks \cite{IEEEexample:funk2009spread,IEEEexample:pei2013spreading,IEEEexample:sun2021competition}. Modelling approaches on multiple processes generally 
employ non-machine learning approaches to model single dynamics on networks, where further research is needed for modelling multiple processes using both non- and data-driven machine learning approaches.

\subsubsection{Interactions of networks and processes}
Network dimension and dynamic process dimension can either be interrelated or independent based on whether one dimension can trigger the dynamics of another dimension to change. Parameter changes of dynamic processes as well as network changes of (i) topology, (ii) attributes and (iii) structural pattern, each indicates the changes of dynamics in the process dimension or the network dimension. Thus, the interrelations exist in two scenarios: (a) certain states of networks trigger the parameter changes of a dynamic process, (b) a dynamic process results in one of the above mentioned three types of the network changes. The interrelations can either be described as a one-way influence that is just about (a) or (b), or a mutual influence that refers to both (a) and (b).

\textbf{Independent VS Interrelated} relations between a network and a dynamic process are discussed considering their changes and the corresponding causes of change within the CNS.

An independent relation between the network and a dynamic process is common in terms of dynamic processes on networks and the research space is dominated by this approach \cite{IEEEexample:zhao2015dynamic,IEEEexample:nowzari2016analysis,IEEEexample:zino2021analysis}. In this case, a dynamic process only influences and causes changes of the node attributes connected with the process itself (e.g. whether a node, as a result of the process, has been infected or adopted new behaviour) rather than a change of the network structure or dynamics. In this scenario parameters of a dynamic process are not altered by the changes in the network.

Networks and dynamic processes with interrelated relations between them consider their mutually or unilaterally triggered changes. Multiple processes that interact with each other via changing network attributes are typical examples of a one-way influence of the type (a), where parameters of spreading dynamics can change with network attributes \cite{IEEEexample:koprulu2019battle,IEEEexample:fu2019analysis,IEEEexample:pan2018effective,IEEEexample:sun2021competition,IEEEexample:funk2009spread,IEEEexample:pei2013spreading,IEEEexample:zhan2018coupling}. For example, in a rumour-truth mixed spreading scenario, the truth spreading rate gets lower when nodes are attributed as rumour-believers \cite{IEEEexample:pan2018effective}. There are also cases for a one-way influence of the type (b), where networks change topology in response to the spread. For example, disease spreading through the network can leave some nodes dead and get them removed \cite{IEEEexample:newman2005threshold}. 
Currently no research is found on interrelation about a mutual influence between the process dimension and the network dimension so the situation where a closed feedback loop between the process and the network is considered.

\textbf{Parameter changes} of dynamic processes are here discussed in terms of interactions and interrelations between networks and processes. All the available examples of processes on networks that are dynamically changing under the impact of networks are about multiple processes, where networks serve as a media for their interactions \cite{IEEEexample:koprulu2019battle,IEEEexample:fu2019analysis,IEEEexample:pan2018effective,IEEEexample:sun2021competition,IEEEexample:koprulu2019battle,IEEEexample:fu2019analysis,IEEEexample:sun2021competition}. Further research is required in the space of dynamic processes that are dynamically changing under the impact of network changes. Network topology that changes in response to the dynamically changing dynamics is another interesting and not addressed research gap.

\subsection{Control mechanisms on CNS}
\label{control}
Control mechanisms of CNS aim to find the optimal strategy to attain its desired state, which involves controllability and synchronization of networks and a control of dynamic processes on networks.

\subsubsection{Network control}
Controllability of networks is achieved using model--based or data--driven approaches \cite{IEEEexample:baggio2021data}. Model--based approaches aim to find an optimal set of driver nodes for CNS under the assumption of a tractable model for network dynamics, where a linear time-invariant systems are often employed to approximate the nonlinear processes that drive the directed networks \cite{IEEEexample:liu2011controllability,IEEEexample:yuan2013exact,IEEEexample:leitold2017controllability}. This approach identifies the driver nodes via maximum matching approach \cite{IEEEexample:lou2020towards}, which enables the calculation of the structural controllability \cite{IEEEexample:liu2011controllability} and exact (state) controllability \cite{IEEEexample:yuan2013exact}. In this context, many studies investigate the impact of topology variations on the controllability of directed networks, ranging from degree distributions \cite{IEEEexample:liu2011controllability}, connection types \cite{IEEEexample:leitold2017controllability}, topology switching \cite{IEEEexample:iudice2019node} to even all possible network structures \cite{IEEEexample:lou2020towards}. To measure the robustness of controllability, simulations of node removal and edge removal attacks are also conducted \cite{IEEEexample:lou2020towards} and convolutional neural networks can be further utilised to improve computation efficiency \cite{IEEEexample:lou2020predicting}. Data--driven based approaches, on the contrary, learn controls from network data without knowing network dynamics \cite{IEEEexample:baggio2021data}. Relevant studies generally focus on undirected, directed or weighted networks, where machine learning methods like reinforcement learning are used to find their optimal control parameters for desired network states \cite{IEEEexample:kiumarsi2017optimal,IEEEexample:baggio2021data}.

Synchronization of CNS, as another type of control towards the desired synchronized state, also involves intensive studies that are well summarised from the perspective of phase oscillator models, stability of synchronised state, and synchronisation in complete or spare networks \cite{IEEEexample:arenas2008synchronization,IEEEexample:dorfler2014synchronization}. In addition, given the similar definitions of synchronization and consensus problem of multi-agent systems \cite{IEEEexample:tang2014synchronization}, they can also be studied from a unified view point by employing ideas about consensus problems across disciplinary areas to complex networks \cite{IEEEexample:fiore2017exploiting,IEEEexample:chen2019synchronization,IEEEexample:lu2019quad}.

\subsubsection{Process control}
Controllability of dynamic processes is achieved via changing networks or introducing another dynamic process.

\textbf{Control via changes of networks} refers to the control of spreading dynamics via changing network topology and attributes, involving non-deep learning--based, deep learning--based and manual strategy--based approaches. Taking control of epidemic spreading processes as an example, researchers seek an optimal set of control actions including topology changes like node removal and edge removal as well as attribute changes via an antidote allocation to minimise infections \cite{IEEEexample:nowzari2016analysis,IEEEexample:kiumarsi2017optimal,IEEEexample:zino2021analysis}. Non-deep learning--based approaches, as is well summarised in \cite{IEEEexample:zino2021analysis}, mimick spreading dynamics, with the non-deep learning models reviewed in section \ref{single}, and optimise the action sequence under corresponding model constraints by a mean-field approximation or geometric programming. Deep learning--based approaches use machine learning methods to seek the optimal action sequence over graphs based on the network information that is incorporated by the network embedding. For example, \cite{IEEEexample:meirom2021controlling} control the epidemic processes over a temporal attributed network using reinforcement learning as a ranking module for actions of changing node attributes, where GNN is encapsulated to embed information about networks and epidemic process. Manual strategy--based approaches focus on the simulation and comparison of manual strategies that change network topology or attributes, like different social network-based distancing strategies proposed and compared to reduce infections of Covid-19 \cite{IEEEexample:block2020social}.

\textbf{Control via interactions between processes} refers to the control of spreading dynamics realised via introducing another spreading dynamics with their competitive, suppressing or supporting impact. For example, the disease containment of a single epidemic spreading dynamics $A$ can be controlled by introducing its competitive process $B$, which is realised by an optimal allocation of a limited number of $B$ spreaders to minimise the spreading of $A$ \cite{IEEEexample:budak2011limiting,IEEEexample:sun2021competition}. The advertising campaign $A$ can also be controlled via introducing its collaborative spreading process $B$, where the best joint advertising campaign can be designed via the optimal allocation of spreader $B$ with the aim of maximising the number of susceptible nodes  \cite{IEEEexample:sun2021competition}.

\section{How do we approach the ultimate goal?}
\label{performance}
In this section, we answer the question of how to approach the ultimate goal of modelling complex networked systems: building a Digital Twin (DT) for real world networked systems. This question is decomposed into three sub-questions: (1) What have we done so far to achieve the goal? (2) How far are we from building the DT for CNS? and (3) How can we move forward?

Existing research on modelling networked systems and their dynamics aims at representing the complex reality through networked structures that minimise information loss and ensure that the model's aims can be fulfilled. The complementary effect of the model's aims fulfilment and minimised information loss contributes to a good model of CNS and its convergence to a DT. To assess the networked systems models and narrow their gaps with DTs, we build an assessment framework from two perspectives: (i) the CNS model's aim fulfilment that diverges when a specific model's aims and an abstract model's aims each focuses on the external tasks and inner rules; and (ii) a DT faithful representation and modelling that helps to merge the requirement of both the specific model's aims about external tasks and the abstract model's aims about its inner rules.

\subsection{What have we done so far?}
\label{whatd}
Researchers have done a lot of work on modelling real world networked systems. A small number of studies have already attempted to develop Digital Twins of complex networked systems for specific application contexts, like IoT systems \cite{IEEEexample:alam2017c2ps,IEEEexample:singh2014survey} and blockchain-encapsulated systems  \cite{IEEEexample:he2018blockchain,IEEEexample:hao2020towards}. However, while many recent studies on modelling, simulation and control of complex networked systems started taking into account the necessary details to faithfully represent aspect of complex reality, none of them explicitly attempted to create a DT of CNS with all its implications which we will now discuss in more detail.

\subsubsection{Bottom-up view: CNS-based attempts}
Researchers have been trying to build a Complex Networked System (CNS) that can faithfully represent and adapt to the real world situation. The networks, as the basic representations of CNS, are approaching reality with more faithful representation of real world information and infusion of evolving dynamics. These attempts partially enable CNS to meet the requirements of Digital Twins in terms of similarity to reality by incorporating complex inner rules and fulfilling external tasks. 

\textbf{Networks} are approaching reality as the structural, temporal, spatial and dynamics dimensions are gradually taken into account. For networks constructed in a data-driven way based on readily observable data set, the description of the the impact of time enables the modelling of data-driven networks to capture the evolving feature of real-world systems, while spatial information represented in the network structure serves as space where dynamics take place. Spatial temporal networks, where temporal networks are modelled under the constrain of space that influences the structure of the networks, are proposed to encompass both temporal and spatial information, which are the closest to the reality data-driven network structures considered until now. For simulation-based networks and hybrid networks, the rule-based simulations have been developed to approach reality as more complex inner rules of complex networked systems are simulated to incorporate the above mentioned complexity dimensions. Networks simulated from microscopic views of agents enable the representation of either real or simulated information in a flexible and faithful manner. From another perspective, the networks have also been widely employed in different scenarios across disciplines, where a wide range of research objects can be represented and analysed with network structures. As is detailed in section \ref{obtain}, networks are employed to represent information in agent-based systems and graph structures, relations that are statistically or semantically extracted and constructed, complex systems infused in networks like IoTs and Blockchains.

\textbf{Dynamics} of and on CNSs with different possible interrelations between those two is another key aspect of building an accurate model of CNS. With the infusion of dynamics into the networks, the CNS is getting further on the road to a DT. Referring to section \ref{quadrant}, there are built-in dynamics that trigger the evolution of networks, while there are also dynamics that take place on the networks under the influence of network structures. CNS composed of networks and dynamics are approaching reality as researchers try to understand and model the interrelations among network dynamics and dynamic processes. Until now, there is considerable literature on dynamic processes on networks that are either static or dynamic. Among different interrelations between dynamics and networks, we find that some studies consider the unilateral interrelation between dynamic processes and static networks. However, we can hardly find any literature on the interrelation of dynamic processes and network dynamics, which is actually the closest to the real world situation.

\subsubsection{Top-down view: DT-directed methods}
The digital twining tasks vary for different cases and require to be adopted and adapted in the context of a CNS. For example, \cite{IEEEexample:jans2020digital} model the DT of an urban-integrated hydroponic farm, where they decompose the modelling process into three crucial elements: data creation that enables an extensive monitoring system for a virtual representation of the farm through data, data analysis that helps to identity key influencing variables, data modelling that enables forecast and feedback. \cite{IEEEexample:kiritsis2011closed} try to shape the actual state and a possible future of the Product Data Technologies from a Closed-Loop Product Lifecycle Management (C-L PLM) perspective, where they see an intelligent product as a product system which contains sensing, memory, data processing, reasoning and communication capabilities at four intelligence levels. \cite{IEEEexample:kapteyn2020probabilistic} view the physical asset and its digital twin as two coupled dynamical systems that evolve over time through their respective state spaces, where the digital twin acquires and assimilates observational data from the asset (e.g., data from sensors or manual inspections) and uses this information to continually update its internal models so that they reflect the evolving physical system. These up-to-date internal models can then be used for analysis, prediction, optimisation and control of the physical system. Referring to the previous studies, we generally decompose the modelling of a DT as tasks including (a) data processing that includes data creation and data integration, (b) data analysis with the purpose of parameter selection, (c) data modelling that enables forecast of eventualities and feedback of real system under the impact of the decision that is made by reference to the forecast.

\textbf{Data processing,} as the fundamental task of modelling a DT, is composed of two parts: (i) data creation enabled by an extensive and robust monitoring system that tracks the observable information, (ii) data integration that features with the record, management and retrieval of information in a real time. In the data creation stage, taking the DT of hydroponic farm built by \cite{IEEEexample:jans2020digital} as an example, they track changing environmental conditions and crop growth through unstructured manual records and a wireless sensor network that sends data in real time to a server. For data integration function, semantic modelling that includes the application of ontology has also been employed to equip the DT with context awareness through record of data, answering queries and information retrieval \cite{IEEEexample:kharlamov2018towards,IEEEexample:boschert2018next,IEEEexample:banerjee2017generating,IEEEexample:hubauer2018use,IEEEexample:lu2020cognitive,IEEEexample:fan2021disaster}. While there also emerges another popular data integration approach that employs blockchain technology. The blockchain can serve as the middleware of IoT with improved interoperability, privacy, security, reliability and scalability \cite{IEEEexample:mandolla2019building,IEEEexample:dai2019blockchain,IEEEexample:azzaoui2021blockchain}. However, in some cases, it is hard to collect and process readily real-time data for a well-established DT in an efficient way and there are studies attempting to deal with problems of data integrity. For example, \cite{IEEEexample:pang2020collaborative} propose a collaborative city digital twin based on federated learning, where multiple city DT can learn a shared model while keeping all the training data locally. This is a promising solution to accumulate the insights from multiple data sources efficiently and avoid violating privacy rules.

\textbf{Data analysis and variable selection} task is of importance to the establishment of DTs, which is presented by \cite{IEEEexample:jans2020digital} as something that includes: (a) the influence of the environment on physical asset, (b) the influence of operable controls on the environment, and (c) the influence of manual changes on the operational controls, where, within the limitations of the data, this exercise identifies the variables which are crucial to track and forecast. Data analysis that enables variable selection is of importance to the modelling of DT. However, most studies ignore this process and just predetermine the observable variables, while the fundamental question of how to identify the minimum number of observable variables has been understudied over the years and needs a systematic research and answer. As is detailed by \cite{IEEEexample:kapteyn2020probabilistic}, a well-designed digital twin should be comprised of models that provide a sufficiently complex digital state space, capturing variation in the physical asset that is relevant for diagnosis, prediction, and decision-making in the application of interest. On the other hand, the digital state space should be simple enough to enable tractable estimation of the digital state, even when only partially observable. Specifically, as a rare example of selecting variables, \cite{IEEEexample:jans2020digital} identify key influencing variables on energy use and crop yield by analysing the relationships between the broad data collected based on temperature, visible radiation, and CO2 levels.

\textbf{Data modelling process} for a DT is essentially a forecasting model that predicts and provides feedback on real world system to help control the DT, which includes two fundamental tasks, a forecast of extended future and a feedback of real-world system. Prediction and inference of reality that has happened before is the basic function of forecasting models, while DTs can further forecast the extended reality by predicting facts that have never happened before. For example, \cite{IEEEexample:chen2017towards} propose a disaster city DT for enhancing disaster response and emergency management processes, where disasters that have never happened before are simulated and real world systems are extensively forecast to enable increased visibility into network dynamics of complex disaster management and humanitarian actions.
The digital twining of complex networked systems are also featured with a decision-making feedback loop with dynamically updated asset specific computational models infused \cite{IEEEexample:kapteyn2020probabilistic}.
Especially in cases of solving multi-objective optimisation problems for complex systems using DT, which are common in analyses of entire product lifecycle in manufacturing, researchers have proposed DT frameworks aimed at multi-objective optimisation with effective feedback from different dynamics. \cite{IEEEexample:cronrath2019enhancing} enhance DTs of autonomous manufacturing systems through reinforcement learning of continuous data fed back from DT, where residual errors between DT and its physical counterpart are compensated and an improved autonomous system can be established. \cite{IEEEexample:zhang2017digital} propose a bi-level iterative coordination mechanism to achieve optimal design performance for AFMS, where an effective feedback of collected decision-support information from the intelligent multi-objective optimisation of the dynamic execution is presented.

\subsection{How far are the CNSs, as they are currently modelled, from DTs?}
\label{howf}
\begin{figure*}[t!]
\centering
\vspace{0.1cm}
\setlength{\abovecaptionskip}{0.3cm}
\setlength{\belowcaptionskip}{-0.3cm}
  \includegraphics[width=6 in]{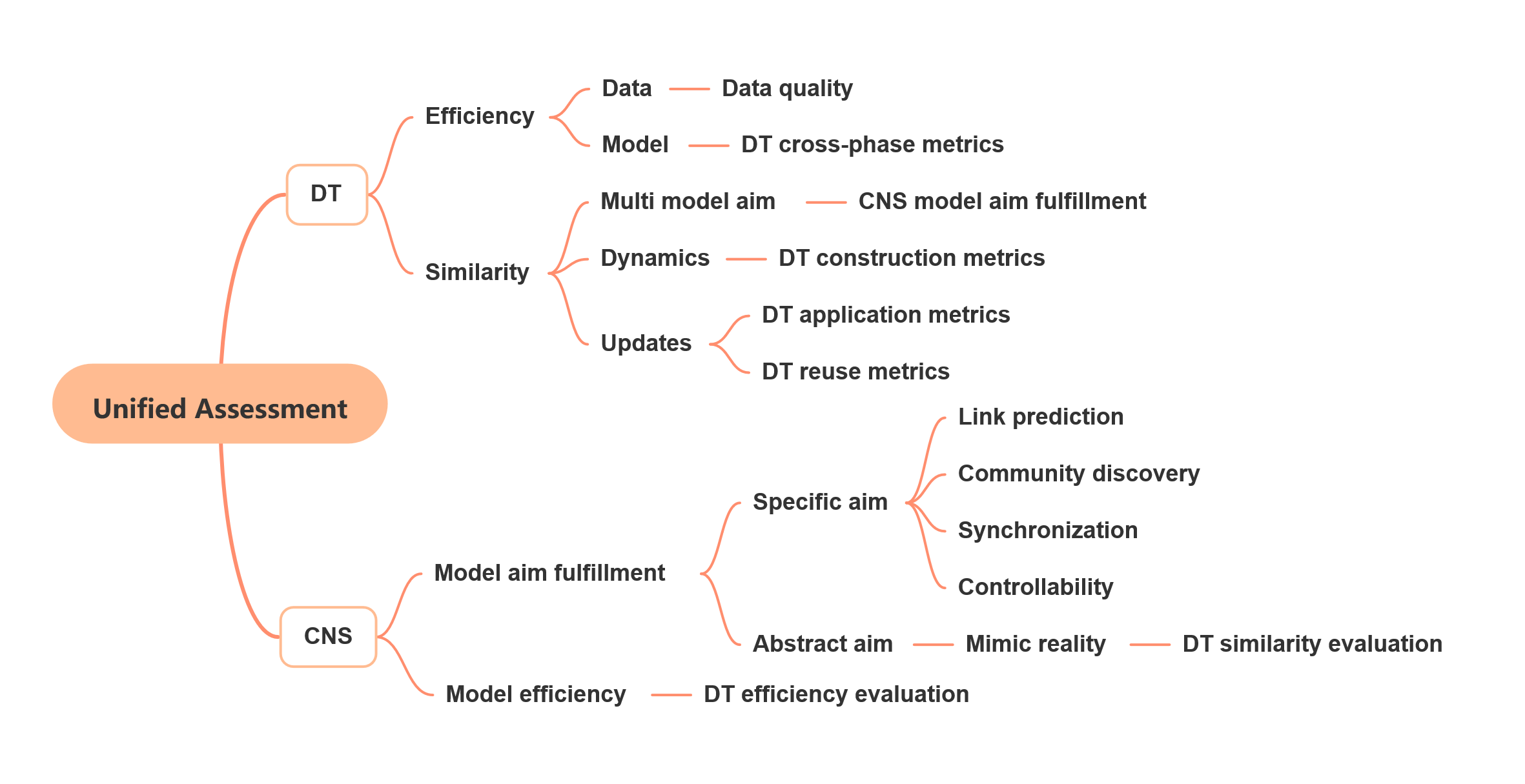}
\caption{The plot of unified assessment criterion for CNS and its distance to a DT}
\label{Assessment}       
\end{figure*}

In most cases, CNS models presented in the literature, while fulfilling the relatively simple model's aims under relatively stringent assumptions, have not been developed with the goal of becoming DTs of their modelled aspects of reality. In effect they only posses partial features of DTs. Therefore, in order to bring the two areas closer together, we propose a unified assessment approach by discussing and attempting to answer the following three questions: (1) What constitutes a good DT? (2) What is a good CNS model? (3) To what extent the current CNS models approach a DT? We try to answer the first two questions with measures that aim at assessing the performance of CNS and metrics that evaluate the quality of DT, as is shown in Fig.\ref{Assessment}, and try to answer question (3) in the context of a good CNS that performs well under certain model's aims and thus has the potential of becoming a DT.


\subsubsection{What constitutes a good DT and how to assess it?}
\label{feature}
DTs are featured with integrated functions like simulation, optimization and data analytics \cite{IEEEexample:ivanov2019digital}. DTs use real-time processing and updates characterised by: (1) real-time connection with the physical entity, (2) self-evolution that enables a DT to learn and adapt in real-time by providing feedback to both the physical asset and the DT, (3) continuous machine learning analysis (dependent on the frequency of the synchronisation), not just a one-time output forecasting, (4) availability of time-series (or time continuous) data for monitoring, (5) level of autonomy that defines if a DT could either make changes to the physical asset itself or if it relies on a human in control who could make changes to the DT, where the property of a DT to be autonomous, not autonomous, or partly autonomous is case-dependent and (6) synchronisation which could be partly continuous or partly event-based \cite{IEEEexample:sharma2020digital}. A good DT should meet the requirement of researchers using relatively simple models while preserving the trust in the data, model and their updates \cite{IEEEexample:wright2020tell,IEEEexample:zhang2021building}. Based on the above features, the assessment of a good DT in the context of CNS can be categorised into two parts: (i) efficiency of data processing and modelling and (ii) similarity with reality from the perspectives of multiple model's aims, self-evolving dynamics and model updates (see Fig.\ref{Assessment}).

The evaluation of efficiency includes data processing efficiency and modelling efficiency. Data processing efficiency involves data quality such as validity and reliability enabled by handling the imperfection of real-time data ranging from imprecision, uncertainty, incompleteness to ambiguity during the process of information retrieval and data integration \cite{IEEEexample:sta2017quality}. Metadata (“data about the data”) captures aspects of the measurement process that may affect the reliability and future usability of the data, which partially addresses trust in data gathered by sensors \cite{IEEEexample:wright2020tell}. The belief function theory is also utilised to estimate the reliability of the information sources \cite{IEEEexample:sta2017quality}. However, as the efficiency of data processing is centered on the observability of experimental physical asset confined by availability of time-series and the real-time connection with the physical entity, it is hard to create quantifiable measures for various evolving application scenarios. For modelling efficiency of a DT, cost, model maturity and model adaptability summarised as a DT cross-phase metrics can be utilised in the assessment \cite{IEEEexample:zhang2021building}, where a high-quality model may cost less in maintenance and reuse. A more mature model gives the expected outcomes and meets application requirements better as time and frequency of using the model increase and a highly-adaptable model recreates the status of the real system better. The cost and model maturity are quantifiable in each application scenario \cite{IEEEexample:crawford2006project,IEEEexample:zhang2007novel}, while the adaptability is hard to be quantified but can be enhanced via parameter sensitivity analysis \cite{IEEEexample:wang2019digital} and continuos monitoring of the model's accuracy over time.

Similarity level between modelled dynamics and reality can be evaluated from the perspectives of multiple model's aims, dynamics and the model updates of parameters in response to the real time data integration and feedback from real systems. A good DT is characterised with multiple model's aim fulfilment and well-handled trade-off between model performance and model complexity. The validation and verification of a good DT is dependent on the model-aim directed evaluation of model output for external tasks and the faithful representation of the inner rules of real systems. In case of a CNS modelled using a DT approach, the evaluation involves the comprehensive application of CNS model-aim evaluation methods (see section \ref{goodCNS}). Similarity of the modelled network dynamics and dynamic processes with that of real world systems can be evaluated using DT construction metrics summarised by \cite{IEEEexample:zhang2021building}, including quantifiable credibility, fidelity, maturity and qualitative description of complexity and DT standardisation, while the similarity in the context of model updates with the evolving reality can also be evaluated using DT application metrics including failure rate and qualitative description of decoupling ability and parallelizabiiity as well as DT reuse metrics including  degree of reconfigurability, reconstructibility and composability. Based on the forementioned methods, the capability of forecasting events that have never happened before and the synchronisation of nodes existing in the CNS can be assessed and enhanced. However, there is no unified quantifiable measures across application scenarios, where the assessment and the comparison of DT modelling methods can be further studied.

\subsubsection{Is it a good CNS?}

\label{goodCNS}
To answer this question, both measures and standards proposed for CNS and DT can be considered towards the goal of building a good DT with high data processing efficiency and similarity of dynamics with the reality. The assessment of CNS can be divided into two parts: (i) model's aim fulfilment assessment, and (ii) model efficiency assessment (see Fig.\ref{Assessment}).

There is considerable literature on the specific model's aims with measurable outputs such as link prediction, community discovery, synchronization, observability and controllability. The evaluation methods for community discovery can be summarised as the internal and external quality evaluation, where more detailed measures can be found in  \cite{IEEEexample:rossetti2018community}. The assessment of synchronization of networks focuses on the stability of identical states \cite{IEEEexample:arenas2008synchronization}, where evaluation of consensus of multi-agent system can also be utilised as they have similar definitions and can be studied from a unified point \cite{IEEEexample:fiore2017exploiting,IEEEexample:chen2019synchronization,IEEEexample:lu2019quad}. The observability, with its dual: controllability, can be categorised as structural and dynamical, each representing observability of topology \cite{IEEEexample:leitold2017controllability}, and variables for coupling nodes and node dynamics \cite{IEEEexample:sendina2019observability,IEEEexample:iudice2019node}. For its evaluation, observability matrix based on the dynamic model for a linear (time-invariant) system proposed by \cite{IEEEexample:kailath1980linear} has been widely used and extended, while the observability and controllability of nonlinear networks are also studied to investigate the effect of nonlinear dynamical interdependences among variables or
the connection with symmetries of networks \cite{IEEEexample:whalen2015observability,IEEEexample:letellier2018nonlinear}.
Particularly, as an evolving model can be mapped to a link prediction algorithm, performance metrics for link prediction can also assist the quantitative comparison of the accuracies of different evolving models \cite{IEEEexample:lu2011link}. For example, link prediction can be utilised to validate dynamic social network simulators with graph convolutional neural networks (GCN) \cite{IEEEexample:ashraf2019simulation}. Link prediction performance can be evaluated using precision\cite{IEEEexample:lu2011link}, Area Under the Precision–Recall (AUPR) curve \cite{IEEEexample:dong2015coupledlp}, Receiver Operating Characteristic (ROC) curves, Area Under the ROC (AUC)\cite{IEEEexample:lu2011link}, Geometric Mean of AUC and PRAUC (GMAUC) \cite{IEEEexample:junuthula2016evaluating}, Error Rate \cite{IEEEexample:chen2018gc}, SumD \cite{IEEEexample:li2014deep}, Kendall's Tau Coefficient (KTC) \cite{IEEEexample:bu2019link} and Micro/Macro/Weighted Average Precision/Recall/F1 Score \cite{IEEEexample:patel2021graph}.

For abstract model's aims such as the mimics of reality using simulation-based networks, we can evaluate these models based on their similarity with the characteristics of reality that they are required to capture, where DT construction metrics \cite{IEEEexample:zhang2021building}, as well as observability and controllability measures for CNS can be utilised. In addition, when the simulation-based networks are encapsulated in complex networked systems with forementioned specific model's aims, the evaluation measures for specific model's aims can also reflect to what extent the network simulations reach the research goals, especially in case of link prediction where similarity of nodes, edges and their dynamics is the focus.

DT feature assessment that is applicable in CNS is mainly about the efficiency of simulation or modelling, as modelling efficiency is a good quality persued by both CNS and DT. DT application metrics and the DT cross-phase metrics utilised in the DT modelling efficiency are also applicable in the context of a CNS, even though it is not a DT. The data processing efficiency can be assessed and persued via observability and controllability measures for CNS, as well as the measures utilised to ensure the trust in the data in the evaluation of data processing efficiency for a DT,

\subsubsection{To what extent is it approaching a DT?}
The assessment of the distance of a CNS model to a DT is built on the prerequisite that the CNS is a good CNS, where a good CNS has partial DT features and has the potential to approach a DT.

\begin{figure}[t!]
\centering
\vspace{0.1cm}
\setlength{\abovecaptionskip}{0.3cm}
\setlength{\belowcaptionskip}{-0.3cm}
  \includegraphics[width=3 in]{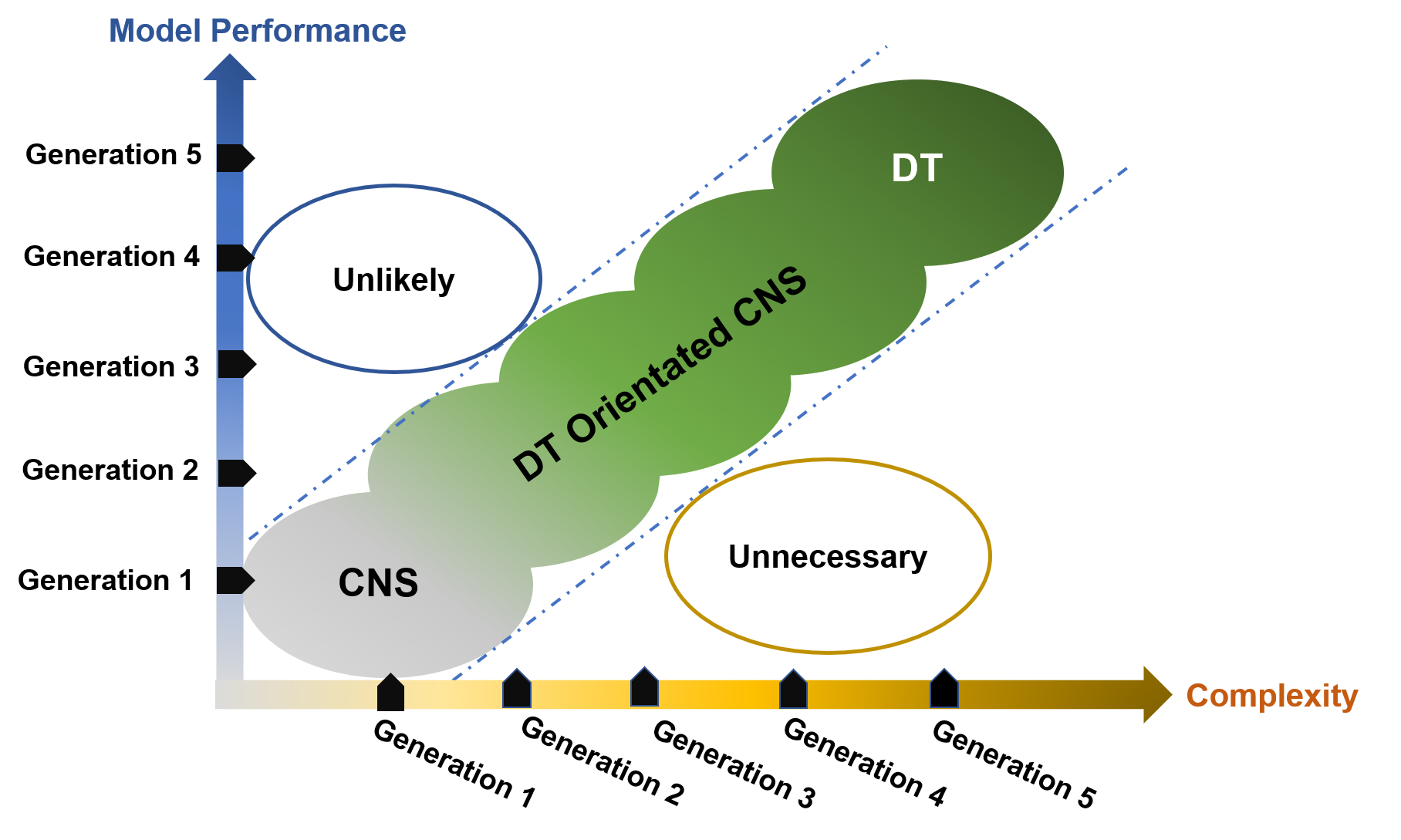}
\caption{The plot of assessing the distance of a CNS to a DT}
\label{Assessmentnew}       
\end{figure}

A DT has an appropriate level of complexity that enables it, with good model performance in terms of faithful representation of real systems, to meet the models' aim. As we have discussed in the previous sections, a CNS can approach a DT with better model performance through appropriately increased complexity. The distance between a CNS and a DT can also be discussed from these two perspectives: (i) complexity and (ii) model performance (See Fig. \ref{Assessmentnew}).


There are no clear boundaries for the development path of a CNS towards a DT, either in terms of model performance or complexity. CNSs with an unnecessary level of complexity, under the lower bound of development path, can be identified when there exists a less complex CNS with equally good or even better model performance. The upper bound of model performance for a CNS naturally exists under the limit of modelling paradigms. When a CNS achieves better model performance through increasing complexity, while falling out of the "unlikely" scope and the "unnecessary" scope, it gets closer to a DT. The bounds of development path for DT-orientated CNS can also be updated with empirical findings in this space.

\begin{figure}[t!]
\centering
\vspace{0.1cm}  
\setlength{\abovecaptionskip}{0.3cm}
\setlength{\belowcaptionskip}{-0.3cm}
  \includegraphics[width=3 in]{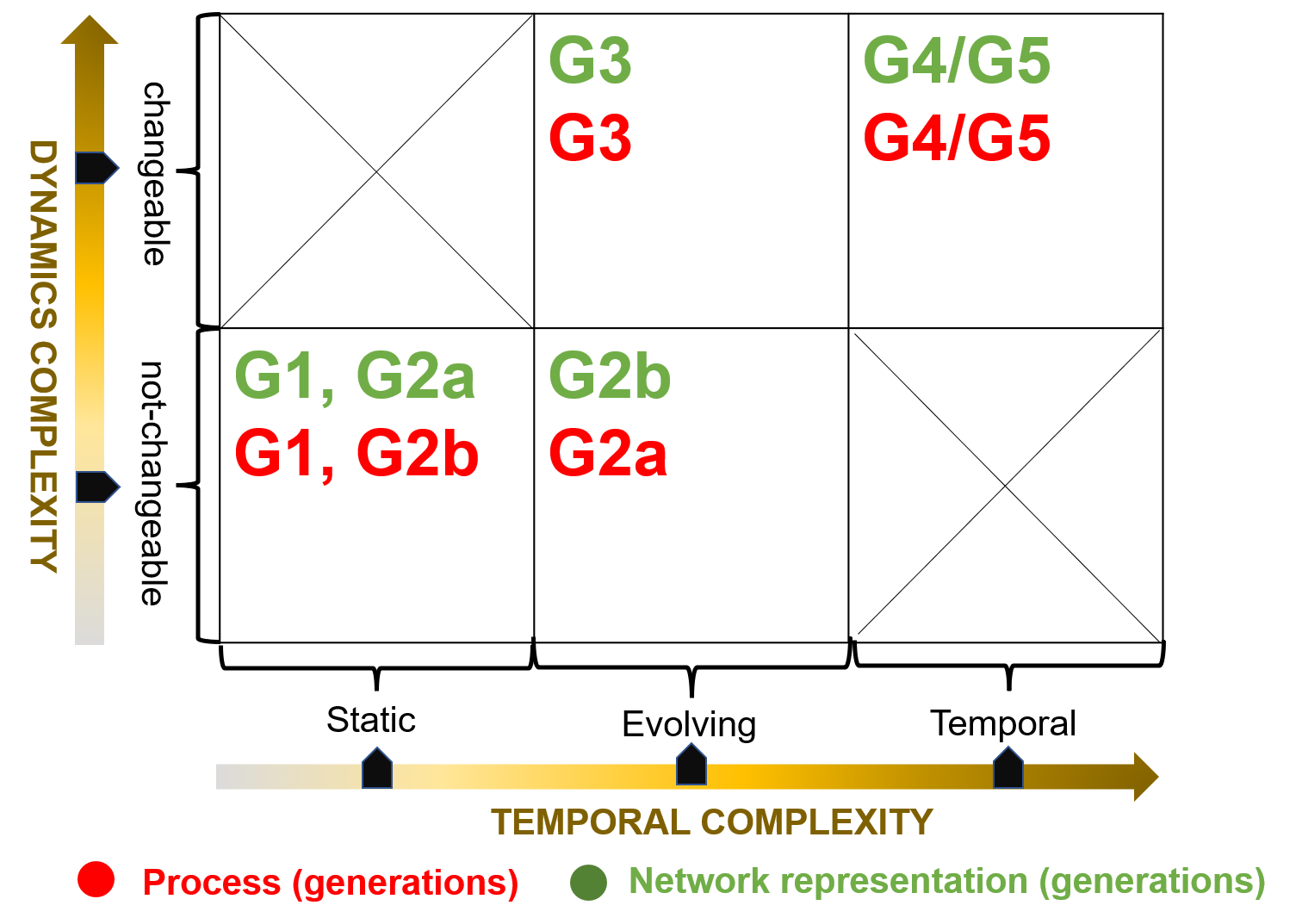}
\caption{The complexity metric of a CNS}
\label{complexityLevel}       
\end{figure}

\paragraph{Complexity} of a CNS is hard to measure using one, concrete measure, but we are able to rank the complexity of (i) network representation in each complexity dimension (see section \ref{NT}), and (ii) CNS modelling based on the 5-generation framework (see section \ref{quadrant}). Based on that, a complexity metric can be identified for each generation of CNSs, including their two components: (i) process and (ii) network representation (see Fig. \ref{complexityLevel}).


For generations of CNSs shown in Fig. \ref{complexityLevel}, their components, the process dimension and the network dimension, are each represented with the coloured symbols of $G_1$, $G_{2a}$, $G_{2b}$, $G_{3}$, $G_{4}$ and $G_{5}$. The CNSs in each generation of models vary in dynamics complexity and temporal complexity, while for structural and spatial complexity, they can be built with any complexity level from those two dimensions. The temporal complexity of a CNS increases as its process, network representation, or both of them, start to change over time in a manner from static (frozen in the time scale), evolving (captured in time windows) to temporal (continuous). The dynamics complexity of a CNS also increases when the dynamics are changeable, with the process modelled based on the changing parameters or the networks evolving via the changes of inner rules. We can compare the complexity of one component of CNSs in one complexity dimension based on this complexity metric. For example, compared with the CNSs in $G_{2a}$, CNSs in $G_{2b}$ are characterised with the temporal complexity that is higher in network representation but lower in process dimension.



\begin{figure}[t!]
\centering
\vspace{0.1cm}  
\setlength{\abovecaptionskip}{0.3cm}
\setlength{\belowcaptionskip}{-0.3cm}
  \includegraphics[width=3 in]{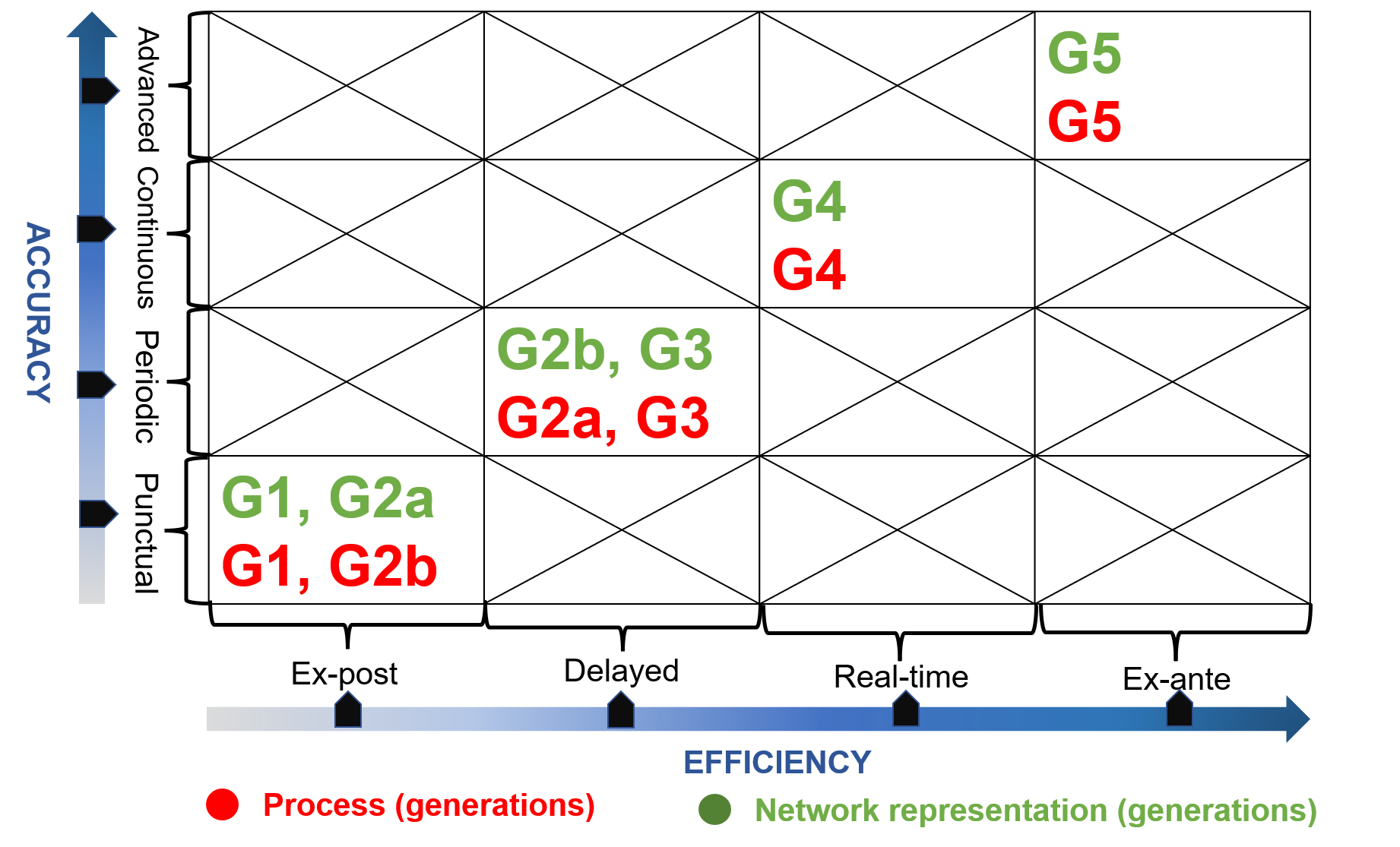}
\caption{The model performance metric of a CNS}
\label{performanceLevel}       
\end{figure}

\paragraph{Model performance} of a CNS can be assessed based on the two requirements of building a good CNS: (i) the model's aim fulfilment, and (ii) the model efficiency (see section \ref{goodCNS} and Fig. \ref{Assessment}). There are both quantitative measures and qualitative description for a CNS assessment, but how to combine them for a comprehensive assessment and to deal with the multi-objective optimisation problem for a good CNS, still requires further study. To gain a rough understanding of the levels of model performance for each generation of CNSs, a model performance metric is built based on (i) the accuracy from the perspective of model's aim fulfilment and (ii) the efficiency (see Fig. \ref{performanceLevel}).


As is shown in Fig. \ref{performanceLevel}, the efficiency is described as ex-post, delayed, real-time and ex-ante based on the CNSs' ways of data processing and modelling. CNSs and their components in the ex-post group have the lowest ranking of efficiency due to the completely post-hoc modelling, like CNSs in $G_1$. CNSs in the delayed group are characterised with streams of snapshots feeding into the systems across time windows with a time lag behind the real systems. CNSs in $G_4$ fall in the real-time group as they conduct real-time data processing and modelling. CNSs in $G_5$, also termed as DTs, are classified in the ex-ante group as they are not only reactive to the observations of real systems in a real-time manner, but also proactive to the things that have never happened before (enabled by the closed feedback loop). The other perspective, i.e. the accuracy, represents a generalised conception of model performance considering how accurately the model' aims are fulfilled. It is classified as punctual, periodic, continuous and advanced. These groups each requires a faithful representation and modelling of the information at only one static time point, within a discrete period, captured continuously or simulated in advance. The required accuracy level increases with the upgraded assessment criterion and the paradigm shift from $G_1$ to $G_5$. For example, the evaluation metrics of community discovery, like the modularity and error rate, are widely used for static networks in $G_1$. They
can be further supplemented with a relative reconstruction error rate to analyse the temporal evolution of dynamic networks and communities in $G_2$ \cite{IEEEexample:wang2016autonomous}. .

\paragraph{Current CNS,} in terms of data efficiency, 
tends to rely more on complex simulation-based networks to be able to capture more realistic features or employ data-driven networks introduced with observable temporal and spatial information. However, the data quality is case-dependent in each application scenario and confined by data sparsity, data security, as well as data processing and representation techniques. It is hard to find studies on CNS built and modelled with real time information because of its observability and the difficulties of building realistic real time data simulator. Though there are studies on CNS built with big data \cite{IEEEexample:bedru2020big}, it is still hard to achieve data efficiency at a "real-time" level. Though in some applications of CNS in a DT like IoT, where networked information can be gathered in a real time by sensors and integrated into a Knowledge Graph or a block chain, such a method is not applicable in all application scenarios given the requirement of equipment. Therefore, there exists a large gap when it comes to data efficiency between a good CNS in current studies and a good DT. On the other hand, for model efficiency, some CNS start to approach a DT with the development of modelling and computation techniques like parallel computation \cite{IEEEexample:cybenko2017parallel}, edge computing \cite{IEEEexample:yang2019integrated,IEEEexample:sun2020makespan} and cloud computing \cite{IEEEexample:ahuja2015trend}, but it is hard to find such empirical research on CNS except for CNS encapsulated in a DT like IoT.

When it comes to the similarity of dynamism of CNS to that of the reality, there are studies on dynamics of the spatio-temporal networks where both time and space are considered to mimic the reality \cite{IEEEexample:zhang2017facial,IEEEexample:xu2017jointly,IEEEexample:morais2021learning}. While there are some studies that model dynamics over dynamic networks with interrelations between dynamics, including interrelations of dynamic processes over networks \cite{IEEEexample:brodka2020interacting} and unilateral influence of dynamic process on static networks \cite{IEEEexample:zhang2021automated}. However, we can hardly find CNS with mutual influence between dynamics on the networks and dynamics of the networks or dynamic processes over spatio-temporal networks with interrelations between dynamics across temporal and spatial dimensions.
Therefore, as the state-of-the-art CNS models have relatively simple aims and often are developed under some strict assumptions, it is hard to find a CNS that fulfills DT standards, unless we focus on DT systems with encapsulated networks like IoT \cite{IEEEexample:yang2014sherlock} and blockchain \cite{IEEEexample:dai2019blockchain}, where the networked systems are built using DT-oriented and directed methods and the network features like topology and interaction of networked node dynamics are ignored.

In terms of model updates, most studies model evolving dynamics in an offline manner without model updates. There are very few attempts to enable model updates e.g. where complex systems are built on networked mobile devices utilising Federated Learning (FL) \cite{IEEEexample:lim2020federated,IEEEexample:abad2020hierarchical}. FL selects random subsets of devices in an offline manner to collect the local model updates and share the updated global model with the devices. This method is also used to deploy distributed data processing and learning in wireless networks in a blockchain encapsulated in a DT \cite{IEEEexample:lu2019blockchain,IEEEexample:lu2020low}. The divergence of model updates remain a future research gap. In addition, the assessment methods of CNS are also in need of updates, as they are required to be more dynamic and able to evolve with model changes.

\subsection{How can we go further?}
\label{further}
To answer the question of how to go further to achieve the ultimate goal of building complex networked systems model that faithfully reflects a real system, we build a framework of CNS-based DT, where the modelling can be decomposed into a series of tasks that require modelling methods from both DT and CNS spaces while setting the half-way point as DT-orientated CNS. We start with listing the research gaps that will guide us in setting goals and tasks for future research.

\subsubsection{Current research gaps}
Based on the five generations modelling framework proposed in section \ref{quadrant} and the conducted review of the state-of-the-art, the current research can realise modelling framework of generation 1 and generation 2 and a small number of approaches can reach generation 3. There is no research on generations 4 and 5, where further studies are required to model this very complex scenarios. To build CNSs under generation 4 or 5, and in this way achieving DT-orientated CNS, seven research gaps need to be tackled:
\begin{enumerate}
    \item fulfilment of external tasks (model aim's) while faithfully mimicking the inner rules of the real system;
    \item meaningful feature extraction and model selection that enables network representation to preserve as much information as needed for the model's aim fulfilment;
    \item network simulation via models built with interpretable inner rules which are able at the same time to deal with structural observability and dynamical observability problems;
    \item dynamics of networks that not only focuses on the topology change but also incorporates attribute changes;
    \item modelling dynamically changing dynamic processes in a way that allows for continuous parameter changes under the impact of network changes;
    \item real time data acquisition and processing for CNS modelling;
    \item the establishment of a feedback loop that enables continuous updates of CNS and the changes of real systems referring to the CNS modelling.
\end{enumerate}

\subsubsection{Set the half-way point: DT-orientated CNS}
DT-orientated CNS emerges with the convergence of DT and CNS modelling approaches, where CNS models approach reality by introducing DT features to the modelling process while a DT incorporates networked information through blockchain, knowledge graph or IoT to assist data processing and modelling. Studies on CNS generally focus on a single model's aim and simple research objectives with predefined set of assumptions. Also, a vast majority of them use historical rather than streaming and continuous data.
DTs can encompass various functions of tools like simulation, optimization and data analytics \cite{IEEEexample:ivanov2019digital} via real-time processing and updates, with research objects ranging from a single product to the society and relaxed assumptions that allow for a CNS representation and modelling in structural, temporal, spatial and dynamics complexity dimensions. When CNS approaches DT more assumptions are being relaxed, more model's aims fulfilled and more complex features can be modelled via more efficient data processing and modelling.

There are still several challenges to overcome and trade-offs to be made on the way to DT-orientated CNS: (i) trade-off between model performance and multiple model's aims, (ii) trade-off between controllability and complexity, (iii) the trade-off between efficiency and accuracy. The existence of multiple model's aims poses a demanding challenge for modelling, which can only be fulfilled by compromising the model performance for a certain model's aim. As complexity of CNS increases with richer information about network components represented and more complex inner rules modelled, it becomes more difficult to control the CNS with limited number of features. To achieve higher accuracy, which is a measure of model performance for external tasks, also requires more complex CNS structures and dynamics, diminishing the efficiency of CNS representation and modelling.

A good DT-orientated CNS refers to the CNS that simulates or models necessary reality to achieve the predetermined model's aim with a DT-level efficiency, which may not need to be a ready DT but is required to deal with the before mentioned trade-offs.
Unified assessment criteria made of mathematical measures for particular model's aims of CNS and the DT evaluation metrics that reflect model efficiency can be used to assess DT-orientated CNS in a dynamic way, as an assistance of model selection and updates. Modelling paradigms of DT across disciplines can also be introduced to CNS to go further on the road to a DT. Therefore, the modelling tasks of DT-oriented CNS and the research gaps are mainly about the resolution of trade-offs between performance (output accuracy, input controllability, model efficiency) and complexity, dynamic assessment of evolving dynamics and the introduction of DT features and paradigms to CNS.

\subsubsection{Embarking on the modelling tasks}
The modelling tasks of DT-orientated CNS should be based on the modelling tasks of CNS and DT that are detailed in section \ref{whatd} and the assessment approach from a unified view point of CNS and DT in section \ref{howf}. The network representation involves tasks of data processing, data analysis and variable selection, while the training of a node and relationships dynamics and dynamics over networks can be summarised as data modelling process, where trust in data, model and updating procedure should be considered with the requirements of observability, similarity and synchronization. Therefore, the modelling tasks of DT-orientated CNS can be generally categorised and presented as:
\begin{enumerate}
    \item Data processing that cope with imperfection of data;
    \item Data analysis and feature selection that considers observability and controllability;
    \item Network representation based on the selected variables and the similarity measures;
    \item Modelling of real-time self-evolving dynamics;
    \item Model updates enabled by reconfiguration and reconstruction;
    \item Model evaluation over the entire process.
\end{enumerate}

For the task (1) on data processing and data management, uncertainty analysis has been utilised to deal with data imperfection, while data fusion emerges as a prevalent way to capture reliable, valuable and accurate information. Also knowledge graphs as well as blockchain have been popular choices for data integration and information retrieval. However, it is challenging to ensure the efficiency of data processing and data management, especially given the "real-time" feature of a DT and the requirements of data quality, where further research is needed. There is also an issue of data sparsity which requires further research on simulation of CNS to deal with the unobservability and unavailability of data.

For the task (2), data analysis and variable selection that considers observability and controllability, has been studied in the context of CNS over the years. However, more effort is needed given the high demand for adaptability of CNS as they approach reality.
Specifically, when it comes to simulation-based networks, how to choose the changeable variables that drive the evolution of networks while preserving the characteristics of real world situation is an interesting research gap given the problem of data scarcity resulting from the rules of keeping data security.

For the task (3), network representation based on the selected variables and the similarity measures, is thoroughly studied research area, while CNS built with DT approaches with the emphasis on network properties has not been deeply studied. Spatio-temporal network, together with the interrelation and interconnection of dynamics within or over such networks are very interesting topics that can be further studied.

The task (4) on modelling of real-time self-evolving dynamics enabled by continuous machine learning is the core element of building a DT, where the interrelations between dynamics in CNS remain an unexplored area especially when it comes to the mutual influence of dynamics on and of networks. More specifically, the evolving dynamics on and of a network is the most relevant to the case-dependent autonomy of DT system, which can be autonomous, not autonomous, or partly autonomous, where the research on the interventions in networked systems can be further introduced with context-awareness and autonomy.

For the task (5), model updates enabled by the reconfiguration and reconstruction, is closely related to the task (4), where the construction of feedback loop is crucial for the continuous modelling process. There is research on a DT built from the perspective of state transition in a discrete way, where how to narrow the time gap between states and extend the state transition to the continuous modelling is a research gap.

For the task (6), model evaluation over the entire modelling process of DT-orientated CNS needs a unified assessment framework, where concrete measures that consider features of both DT and CNS should be explored. There is already literature on measures of network analysis and DT analysis, though the principled combination of these two or proposing new integrated quality measures remain an outstanding research gap.

The above tasks show the complexity of the research that is needed to be accomplished along the way of working towards DT-orientated Complex Networked Systems.

\section{Conclusions}
This survey focuses on the modelling approaches of Complex Networked Systems that pave the path for its convergence to the ultimate goal: a Digital Twin of a CNS.

We review and discuss the CNS from three perspectives: (i) model's aims that have been studied for CNS (see section \ref{section2}), (ii) modelling paradigms that enable to represent a networked system in a way that preserves as much information as needed (see section \ref{how1}), (iii) modelling approaches for dynamics of networks and dynamics over networks that enable to meet model's aims (see section \ref{quadrant}). Those themes are discussed through the lenses of four, proposed by us, complexity dimensions of complex systems: (i) structural, (ii) temporal, (iii) dynamics and (iv) spatial. A discussion that considers those complexity dimensions  enables to better understand current modelling challenges and quantify how far we are from achieving Digital Twin modelling capabilities when representing networked systems.

The model's aims for CNS distinguish when they focus either on the external task to be performed on the system or modelling of the inner rules of real systems, and this division and specialisation can be eliminated when a Digital Twin is considered since it is able to undertake multiple external tasks by faithfully covering and reflecting the complexities of real systems. Models of CNS proposed over the years have been found to approach real systems with increasing complexity in structural, temporal, spatial and dynamics dimensions. To generate and preserve this heterogeneous networked information, modelling paradigms of network representation get more complex with compromised interpretability. These models either focus on inner rules of network generation at a local level or aim at a compressed network representation at a global level, but all converge to the goal for a faithful representation of real systems.

Dynamics of networks, dynamic processes on networks as well as their interrelations are three elementary sources of complexity for dynamics in the CNS. To navigate a pathway through different levels of complexity of modelling CNS, we devise a modelling framework of CNS that considers all these three elements and consists of five generations reflecting the progress of work that has been done in this field. Each generation builds upon the previous one meaning that the next generation encompasses higher complexity levels than the previous one.

This modelling framework is agnostic to the model's aim so any of the discussed aims can be attempted using models built within each of the generations. Though, one needs to remember that models from different generations will enable to achieve selected aim to a different extent. The proposed framework also shows how Complex Networked Systems' models approach a Digital Twin with more complexity through generations, (i) generation 1: dynamic process on static networks, (ii) generation 2 with two variations: dynamic process on evolving networks and evolving dynamic process on static networks, (iii) generation 3: evolving dynamic processes on evolving networks with interrelations between them, (iv) generation 4: temporal dynamic processes on temporal networks with interrelations between them and the acquisition of real time information and finally (v) generation 5 that further introduces modelling framework of generation 4 with an information feedback from CNS's model to the real system. From generation 1 to generation 5, the real system can be represented more faithfully with richer information captured and finally a CNS-based DT can be created in generation 5. Current studies have made good progress under modelling framework of generation 1 and generation 2. Only small number of approaches reach generation 3. For generation 4 and 5, there is no research in this space and further studies are required to model this very complex scenarios to achieve better performance in the context of any of the presented model's aims.

To be more aware of how to approach a Digital Twin with CNS, we propose an assessment framework (see section \ref{performance}) that aims at quantifying the distance of CNS to DT from the perspective of CNS model's aim fulfilment and the perspective of a DT's faithful representation of reality. A half-way point referred to as a DT-orientated CNS is proposed to bridge the gap between the current approaches to modelling of CNS and the ultimate goal of a DT (generation 5 models) for future study.

The goal of the future research in the space of complex networked systems and network science more broadly is to develop a DT-orientated CNS that will enable to address research gaps presented in Section \ref{further}. Integrating dynamic networks with dynamic processes and allowing for mutual influence between them (allowing at the same time for continuous adaptation of the system using streaming data as an input) will make it possible to create the DT-orientated CNSs. This will be a major breakthrough in the space of modelling CNS.

\section*{Acknowledgements}\label{section-acknowledge}
This work was supported by the Australian Research Council, Dynamics and Control of Complex Social Networks under Grant DP190101087.



\end{document}